\documentclass[aps,prl,twocolumn,superscriptaddress,nobibnotes]{revtex4-2}

\usepackage{amsmath,amsfonts,amssymb,amsthm,mathtools}

\usepackage[english]{babel}
\usepackage[utf8]{inputenc}
\usepackage[T1]{fontenc}

\usepackage{latexsym}
\usepackage[pdftex]{graphicx}
\usepackage{epstopdf}

\usepackage{subfigure}
\usepackage{enumitem}

\usepackage{hyperref}
\usepackage[all]{hypcap}

\usepackage{fancyhdr}

\fancypagestyle{plain}{
    \fancyhf{}
    \fancyfoot[C]{\thepage}
    
}

\hypersetup{
  pdftitle={},
  pdfauthor={},
  colorlinks=true,
  linkcolor=blue,
  citecolor=blue,
  filecolor=black,
  urlcolor=blue,
  bookmarksopen,
  bookmarksopenlevel=1,
}

\newcommand{\cA}{\mathcal{A}}

\newcommand{\cD}{\mathcal{D}}

\newcommand{\cF}{\mathcal{F}}

\newcommand{\cO}{\mathcal{O}}
\newcommand{\cP}{\mathcal{P}}

\newcommand{\mfu}{\mathfrak{u}}

\newcommand{\ee}{\mathrm{e}}
\newcommand{\ii}{\mathrm{i}}
\newcommand{\dd}{\mathrm{d}}

\def\diag{\operatorname{diag}}

\def\wDiff{\widetilde{\operatorname{Diff}}}

\def\hc{\mathrm{h.c.}}

\def\kF{k_{\textrm{F}}}

\def\uv{\underline{v}}
\def\uf{\underline{f}}
\def\ux{\underline{x}}

\newcommand{\pdag}{^{\vphantom{\dagger}}}

\newcommand{\ppr}{^{\vphantom{\prime}}}

\newcommand{\wick}[1]{\left. :\! \hspace{-0.5pt} #1 \hspace{-0.5pt} \!: \right.}

\renewcommand{\Re}{\operatorname{Re}}

\DeclareMathSymbol{\shortminus}{\mathbin}{AMSa}{"39}

\theoremstyle{definition}

\theoremstyle{remark}

\begin{document}

\title{%
Perfect Wave Transfer in Continuous Quantum Systems%
}%

\author{Per Moosavi}
\email{per.moosavi@fysik.su.se}
\affiliation{%
Department of Physics, Stockholm University, 10691 Stockholm, Sweden%
}%

\author{Matthias Christandl}
\email{christandl@math.ku.dk}
\affiliation{%
\mbox{Department of Mathematical Sciences, University of Copenhagen, Universitetsparken 5, 2100 Copenhagen, Denmark}%
}%

\author{Gian Michele Graf}
\email{gmgraf@phys.ethz.ch}
\affiliation{%
Institute for Theoretical Physics, ETH Zurich,
Wolfgang-Pauli-Strasse 27, 8093 Z{\"u}rich, Switzerland%
}%

\author{Spyros Sotiriadis\vspace{-0.8mm}}
\email{spyros.sotiriadis@physics.uoc.gr}
\affiliation{%
\mbox{Institute of Theoretical and Computational Physics, Department of Physics, University of Crete, 71003 Heraklion, Greece}%
}%
\affiliation{%
\mbox{Dahlem Center for Complex Quantum Systems, Freie Universit\"{a}t Berlin, 14195 Berlin, Germany}%
}%

\date{%
February 22, 2026%
}%

\begin{abstract}
The transfer of information from one part of a quantum system to another is fundamental to the understanding and design of quantum information processing devices. In the realm of discrete systems such as spin chains, inhomogeneous networks have been engineered that allow for the perfect transfer of qubits from one end to the other. Here, by contrast, we investigate the perfect transfer of information in continuous systems, phrased in terms of wave propagation. A remarkable difference is found between systems that possess conformal invariance and those that do not. Systems in the first class enjoy perfect wave transfer (PWT), explicitly shown for one-particle excitations and anticipated in general. In the second class, those that exhibit PWT are characterized as solutions to an inverse spectral problem. As a concrete example, we demonstrate how to formulate and solve this problem for a prototypical class of bosonic theories, showing the importance of conformal invariance for these theories to enjoy PWT. Using bosonization, our continuum results extend to theories with interactions, broadening the scope of perfect information transfer to more general quantum systems.
\end{abstract}

\maketitle

\thispagestyle{fancy}
\pagestyle{fancy}

\setlength{\skip\footins}{1.2pc plus 5pt minus 1pt}

\setlength{\abovedisplayskip}{7pt}
\setlength{\belowdisplayskip}{7pt}

\emph{Introduction}---%
\csname phantomsection\endcsname%
\addcontentsline{toc}{section}{Introduction}%
%
A fundamental task in quantum information processing is to transfer information from one register to another \cite{NikolopoulosJex:2014, NokkalaEtA:2024}.
Optical fibers may be used, but efficiency can be improved using a quantum wire as both communication channel and registers \cite{Bose:2003}.
Since this relies on collective phenomena to transfer quantum states, efficiency is optimized by engineering the system's many-body properties.
This allows not only for high-fidelity but even perfect transfer of states without dynamical control of individual sites.
Such \emph{perfect state transfer} (PST) was first shown theoretically for wires modeled as spin chains with inhomogeneous hopping \cite{ChristandlEtAl:2004, AlbaneseEtAl:2004}, then extended to spins on graphs and other networks \cite{ChristandlEtAl:2005, BoseEtAl:2005, Bose:2007, Kay:2010, VinetZhedanov:2012, Godsil:2012, KemptonEtAl:2017b, MeiEtAl:2018, DerevyaginEtAl:2020, Huang:2022} and demonstrated experimentally \cite{ZhangEtAl:2005, Perez-LeijaEtAl:2013, ChapmanEtAl:2016, LiEtAl:2018, XiangEtAl:2024, RoyEtAl:2024}.
However, PST has so far been addressed only in the discrete realm, while powerful continuum frameworks to describe collective quantum phenomena remain untapped.

\begin{figure}[t]
\centering
\includegraphics[scale=0.40]{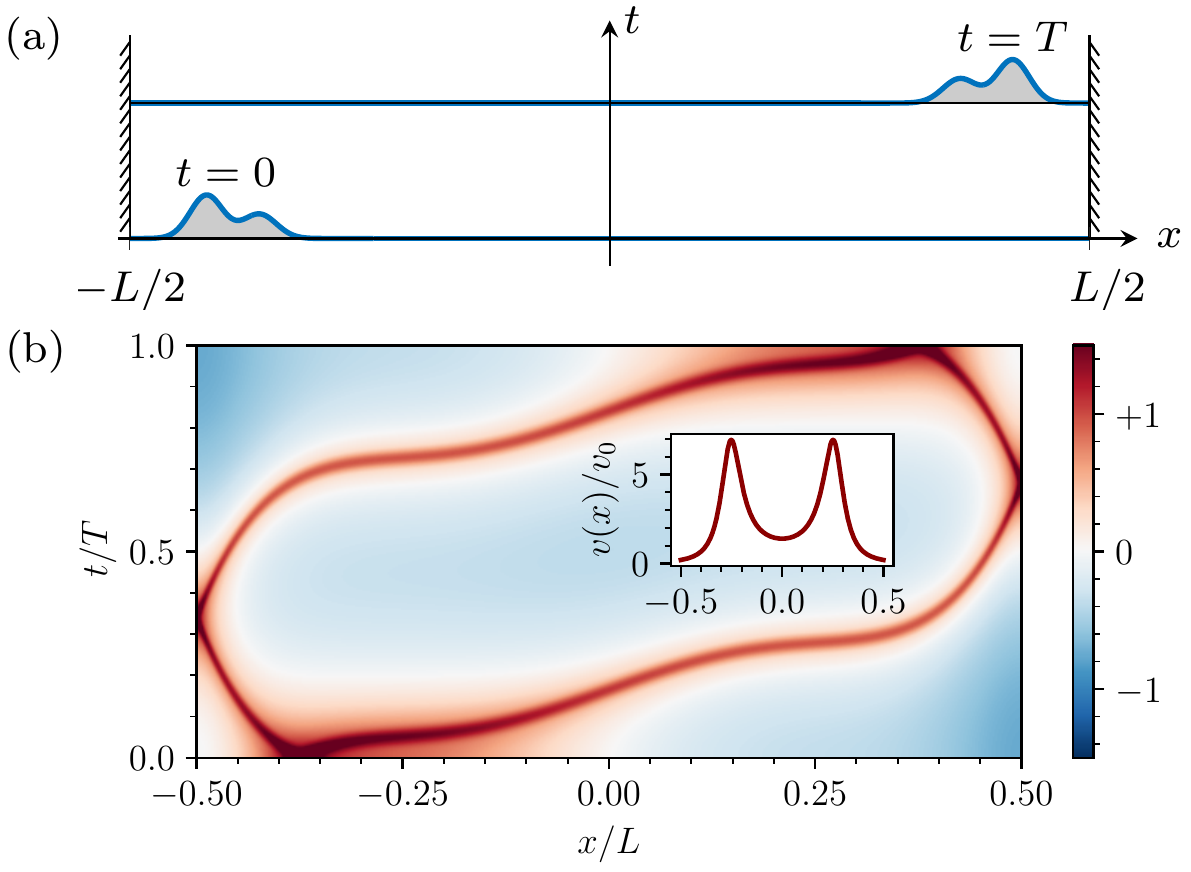}
\vspace{-2mm}
\caption{%
(a) Illustration of PWT with Neumann BCs showing an initial wave that evolves to its reflection at a certain time $T$.
(b) Its manifestation in an inhomogeneous CFT with a given profile $v(x)/v_{0}$ (inset), showing a localized initial wave propagating along curved light-cone trajectories that recombine at the reflected point at $t = T \equiv L/v_{0}$ given by Eq.~\eqref{yx_v0_def}.
Specifically, the real part of boson two-point correlations \eqref{varphivarphi} at the free-fermion point is plotted, which by bosonization encodes the same information as $\log F(t)$ in Eq.~\eqref{Ft} for deltalike waves $\xi_{1}$ centered at $-3L/8$ and $\xi_{2}$ at $x$ (setting $\kF = 0$ for simplicity and $t' = 0$).%
}%
\label{Fig:MCGS_fig1}
\vspace{-3mm}
\end{figure}

In this Letter, we harness that power to investigate quantum information transfer in inhomogeneous continuous systems in 1+1 dimensions (1+1D).
At low energies, such examples are realizable using, e.g., spin ladders \cite{KlanjsekEtAl:2008}, Josephson-junction arrays \cite{CedergrenEtAl:2017}, ultracold atoms \cite{YangEtAl:2017}, and nanowires \cite{CCGOR:2011, MistakidisEtAl:2023, BouchouleEtAl:2025}.
In this framework, rephrasing using waves becomes natural, motivating the term \emph{perfect wave transfer} (PWT): As for states in PST, a system on an interval $[-L/2, L/2]$ of length $L$ with suitable boundary conditions (BCs) exhibits PWT if propagating any initial wave for some time $T$ amounts to a reflection; see Fig.~\ref{Fig:MCGS_fig1}.
Concretely, \emph{there is PWT} if
\begin{equation}
\label{PWT_def}
\langle \cO(x, 0) \rangle = \langle \cO(-x, T) \rangle
\end{equation}
for all time-evolved observables $\cO(x, t)$ and expectation values $\langle \cdot \rangle$ with respect to any state.
The dynamics will be time independent, but the problem is also interesting for driven systems \cite{LiEtAl:2018, XiangEtAl:2024, RoyEtAl:2024}, with randomness \cite{YaoEtAl:2012}, or relaxed to a property for specific classes of observables or states.
Ultimately, our motivation for understanding PWT is the efficient high-fidelity transfer of information---a problem that has attracted recent attention, e.g., in the design of distributed networks for quantum computation \cite{MagnardEtAlWallraff:2020}.

Related advances include nonequilibrium studies of inhomogeneous versions of conformal field theory (CFT), i.e., conformally invariant quantum field theories \cite{FMS:1997} deformed so that the usual constant propagation velocity becomes a profile $v(x)$ \cite{ADSV:2016, DSVC:2017, DSC:2017, GLM:2018, LaMo2:2019, Moosavi:iCFT:2024}.
Among their applications, such \emph{inhomogeneous CFTs} effectively describe gapless quantum many-body systems with mesoscopic inhomogeneities, e.g., spin chains with spatially varying hopping and trapped ultracold atoms; cf.\ \cite{GMS:2022}.
Special cases include the profile $v(x) = v \sqrt{1 - (2x/L)^2}$ for some constant $v > 0$ \cite{WenRyuLudwig:2016, LiuEtAl:2025}---the continuum counterpart of the inhomogeneous spin chain \cite{ChristandlEtAl:2004} first exhibiting PST---and profiles produced by M\"{o}bius transformations \cite{Katsura:2012, WenWu:2018, MacCormackEtAl:2019,  LapierreEtAl:2020, FanEtAl:2020, DasEtAl:2021, CaputaGe:2023, GotoEtAl:2024}.
Note that for PWT, as for other transport properties in finite systems, the choice of BCs \cite{GawedzkiKozlowski:2020, TajikEtAl:2023, LiuEtAl:2025, BernamontiEtAl:2024} is important.

Engineering PWT will, at heart, be a spectral problem, amounting to whether the propagator coincides with the parity operator at certain times.
This is shown for prototypical quantum field theories describing an inhomogeneous quantum wire---our continuous channel---focusing on universal low-energy descriptions where conformal invariance can be parametrically turned on or off.
Notably, these are directly applicable to ultracold atoms \cite{RauerEtAl:2018, GluzaEtAl:2022, TajikEtAl:2023}, while such experimental systems are even fundamentally continuous \cite{BDZ:2008}, in itself warranting a continuum treatment of information transfer.
Remarkably, for bosonic theories \cite{Tomonaga:1950, Luttinger:1963, MattisLieb:1965, Haldane:1981, MaSt:1995, SaSc:1995, Pono:1995, Moosavi:DBdG_iQLs:2023}, we arrive at an inverse spectral problem for Sturm-Liouville operators.
Its solution determines if the channel enjoys PWT, and we show that conformal invariance is important if the system is required to be sufficiently regular.
Using bosonization \cite{Voit:1995, DelftSchoeller:1998, SchulzCunibertiPieri:2000, Giamarchi:2003, Cazalilla:2004, LaMo1:2015}, our results extend PWT or PST to a large class of theories, including interacting fermions, and we also discuss a proposal to extend beyond low energies.
Lastly, in addition to quantum communication in solid-state devices \cite{Lloyd:2008, ZengEtAl:2019}, our PWT results relate to relativistic quantum information \cite{Fuentes-SchullerMann:2005, MartinMartinezEtAl:2011}, and potentially to communication over long distances \cite{Kimble:2008}.

\emph{Inhomogeneous CFT}---%
\csname phantomsection\endcsname%
\addcontentsline{toc}{section}{Inhomogeneous CFT}%
%
To begin, we study the continuum limit of inhomogeneous spin chains \cite{Bose:2003, ChristandlEtAl:2004}.
In the gapless regime, their continuous counterpart is an inhomogeneous CFT with a velocity profile $v(x) > 0$ \cite{ADSV:2016, DSVC:2017, DSC:2017, GLM:2018, LaMo2:2019, Moosavi:iCFT:2024}.
The simplest example is that of inhomogeneous free fermions, given by the Hamiltonian (setting $\hbar = 1$) \cite{Note:Casimir}
\begin{equation}
\label{H_iFF}
H_{\textrm{iFF}}
\equiv
\int_{-L/2}^{L/2} \hspace{-2.3pt}
\dd x\, \frac{v(x)}{2}
\Bigl[ \!
  \wick{ \psi^\dagger_{+} (-\ii) \partial_x \psi\pdag_{+} }
  + \wick{ \psi^\dagger_{-} \ii \partial_x \psi\pdag_{-} }
  + \hc
\Bigr]
\end{equation}
with fields $\psi_{\pm}(x)$ of right- ($+$) and left- ($-$) moving fermions satisfying $\{ \psi\pdag_{\pm}(x), \psi^{\dagger}_{\pm}(x') \} = \delta(x-x')$ and $\{ \psi\pdag_{\pm}(x), \psi^{(\dagger)}_{\mp}(x') \} = 0$.
Here, $\wick{ \cdots }$ denotes Wick (normal) ordering with respect to the ground state $|\Omega\rangle$ (which, for bilinears in the fields, amounts to an additive renormalization).
For simplicity, we focus on Eq.~\eqref{H_iFF}, while general CFTs are discussed in the Supplemental Material (SM) \cite{SM}.
We assume open (Neumann) BCs at $x = \pm L/2$, incorporated by standard unfolding methods; cf.\ \cite{EggertAffleck:1992, FabrizioGogolin:1995, Giamarchi:2003, Cazalilla:2004, GawedzkiKozlowski:2020, ChuaEtAl:2020}:
Couple right and left movers by identifying $\psi_{-}(x) = \psi_{+}(L-x)$, giving a chiral theory of only right-moving fermions on $[-L/2, 3L/2]$ with periodic BCs and velocity $v(x)$ for $x \in [-L/2, L/2]$ and $v(L-x)$ for $x \in [L/2, 3L/2]$ \cite{Note:Particle_flow}.

PST was probed in \cite{ChristandlEtAl:2004} using a time-dependent overlap between one-particle states that were mutual reflections.
Their continuum counterparts are states created by fields
\begin{equation}
\label{psi_def}
\psi(x)
\equiv
\ee^{\ii \kF x} \psi_{+}(x) + \ee^{-\ii \kF x} \psi_{-}(x),
\end{equation}
where $\kF$ denotes the Fermi momentum \cite{Note:Eq_psi_def}.
As a subset of possible states, consider
\begin{equation}
\label{Psi_j}
|\Psi_{j} \rangle
\equiv
\int_{-L/2}^{L/2} \dd x\, \xi_{j}(x) \psi^{\dagger}(x) |\Omega\rangle
\qquad (j = 1,2)
\end{equation}
for smooth enough functions $\xi_{j}(x)$, interpreted as ``waves'' of fermionic excitations in each state, and study the time-dependent overlap
\begin{equation}
\label{Ft}
F(t)
\equiv
\langle\Psi_{2}| \ee^{-\ii H_{\textrm{iFF}} t} |\Psi_{1}\rangle.
\end{equation}
The states in Eq.~\eqref{Psi_j} form a reduced class, as only one-particle excitations are possible, and right and left movers are equally distributed.
(The latter restriction is lifted in the SM \cite{SM}.)

To probe PWT for one-particle excitations as in \cite{ChristandlEtAl:2004}, we restrict to specular waves, $\xi_{2}(x) = \xi_{1}(-x)$, and compare $F(t)$ in Eq.~\eqref{Ft} with $\langle\Psi_{1}|\Psi_{1}\rangle$.
Specifically, \emph{there is one-particle PWT} if there exists $T > 0$ so that
\begin{equation}
\label{PWT_op_def}
|F(T)|
= |\langle\Psi_{1}|\Psi_{1}\rangle|
\end{equation}
for all initial waves $\xi_{1}(x)$ in Eq.~\eqref{Psi_j} \cite{Note:PWT_op_def}.

Eq.~\eqref{PWT_op_def} amounts to a comparison between two-point correlation functions at equal and nonequal times.
Such correlations can be computed in any inhomogeneous CFT by mapping them to the corresponding homogeneous correlations \cite{Moosavi:iCFT:2024}.
The details given in the SM \cite{SM} distill to comparing spacetime points after conformal transformations to coordinates
\begin{equation}
\label{yx_v0_def}
y = y(x) \equiv \int_{0}^{x} \dd s\, \frac{v_{0}}{v(s)},
\qquad
\frac{1}{v_{0}}
\equiv \frac{1}{L} \int_{-L/2}^{L/2} \frac{\dd s}{v(s)}.
\end{equation}
The comparison's outcome shows that reflection symmetry is essential:
\emph{Inhomogeneous CFTs exhibit one-particle PWT if and only if $v(x)$ is even.}
Moreover, this transfer occurs at odd multiples $(2m+1)T$ (for $m = 0, 1, \ldots$) of the time $T = L/v_{0}$ for excitations to cross the system's entire length.
Note that $y \in [-L/2, L/2]$ if $v(x)$ is even.

As an example, consider the square-root profile $v(x) = v \sqrt{1 - (2x/L)^2}$, effectively describing the spin chain in \cite{ChristandlEtAl:2004}.
It exhibits PWT, just like for any even profile $v(x)$, in contrast to \cite{ChristandlEtAl:2004}, where that profile was exceptional in this respect.
Notably, the square-root profile is positive everywhere except at $x = \pm L/2$.
Thus, strictly speaking, $v(x)$ should be regularized before using the tools in \cite{Moosavi:iCFT:2024}, but since the integral defining $1/v_{0}$ in Eq.~\eqref{yx_v0_def} converges, it has no consequence for PWT \cite{Note:Limit_Mobius}.

\emph{Bosonic theories}---%
\csname phantomsection\endcsname%
\addcontentsline{toc}{section}{Bosonic theories}%
%
The above result and observations should not come as a surprise.
Indeed, if one compares discrete and continuous theories, then an inhomogeneous CFT can describe only low-energy excitations of a given lattice model.
Moreover, we show \cite{SM} that any Hamiltonian (with pure point spectrum) must commute with the parity operator for PWT to be possible, i.e., $v(x)$ must be even, as should be expected.
Furthermore, since excitations propagate ballistically in CFT, one might anticipate that this is not only necessary but also sufficient to ensure PWT with respect to all states.
E.g., the model \eqref{H_iFF} has a linear spectrum and is free, suggesting that PWT for one-particle excitations extends to many particles.
The rest of this Letter concerns to what extent this is true more generally:
What are \emph{necessary} and \emph{sufficient} conditions for PWT as defined in Eq.~\eqref{PWT_def}?

To answer this question and understand the role of conformal invariance, we should \emph{not} assume the latter.
Specific nonconformal examples include 1+1D fermions with nonlocal interactions \cite{MastropietroWang:2015, LLMM1:2017, LLMM2:2017} or the Lieb-Liniger model \cite{LiebLiniger:1963}.
While interesting in their own right, we address this question using a paradigmatic low-energy theory, namely, inhomogeneous Tomonaga-Luttinger liquids (TLLs) \cite{Tomonaga:1950, Luttinger:1963, MattisLieb:1965, Haldane:1981, MaSt:1995, SaSc:1995, Pono:1995, Moosavi:DBdG_iQLs:2023}.
By bosonization, this covers fermionic models generalizing the one in Eq.~\eqref{H_iFF}, while manipulating the inhomogeneities allows one to controllably turn conformal invariance on or off.
The theory is formulated using a compactified bosonic field $\varphi(x)$ (modulo $2\pi$) with conjugate $\Pi(x)$ satisfying $[\partial_x \varphi(x), \Pi(x')] = \ii \partial_x \delta(x-x')$.
The Hamiltonian is \cite{Note:Casimir}
\begin{equation}
\label{H_iTLL}
H_{\textrm{iTLL}}
\equiv
\int_{-L/2}^{L/2} \frac{\dd x}{2\pi} v(x)
  \! \wick{
    \biggl(
      \frac{[\pi \Pi(x)]^2}{K(x)}
      + K(x)[\partial_x \varphi(x)]^2
    \biggr)
  } \! ,
\end{equation}
where, in addition to $v(x) > 0$, the Luttinger parameter $K(x) > 0$ also depends on $x$.
For simplicity, we assume that both functions are smooth.
For $K(x) = K$ constant, Eq.~\eqref{H_iTLL} is an inhomogeneous CFT, and if $K = 1$, it is the bosonized version of the free-fermion model \eqref{H_iFF} \cite{Voit:1995, DelftSchoeller:1998, SchulzCunibertiPieri:2000, Giamarchi:2003, Cazalilla:2004, LaMo1:2015}.
However, $K'(x) \neq 0$ locally couples right and left movers nontrivially; thus, they do not commute, and the theory is not conformal \cite{Moosavi:DBdG_iQLs:2023}.
Again, we impose open BCs:
\begin{equation}
\label{jmath_BCs}
\hspace{-1mm}
\jmath(x) \big|_{x = \pm L/2} = 0,
\;\;
\textnormal{where}
\;\;
\jmath(x)
\equiv
- \frac{v(x)K(x)}{\pi} \partial_x \varphi(x)
\end{equation}
is the particle current.
As a rule, we further assume that $K'(x)$ is zero at $x = \pm L/2$ so that right and left movers are coupled only via the BCs \eqref{jmath_BCs} at those points \cite{Note:BCs_K(x)}.

\emph{Sturm-Liouville (SL) theory}---%
\csname phantomsection\endcsname%
\addcontentsline{toc}{section}{Sturm-Liouville (SL) theory}%
%
Following \cite{Stringari:1996, HoMa:1999, MenottiStringari:2002, Ghosh:2004, PetrovEtAl:2004, CitroEtAl:2008, GMS:2022}, we use SL theory to find eigenfunctions, using which the fields can be expanded and the Hamiltonian \eqref{H_iTLL} diagonalized \cite{Note:GMS_vs_DBdG}.
Indeed, using integration by parts and Eq.~\eqref{jmath_BCs},
\begin{equation}
\label{H_iTLL_cA}
H_{\textrm{iTLL}}
=
\int_{-L/2}^{L/2} \frac{\dd x}{2\pi} w(x)
  \! \wick{
    \Biggl(
      \biggl[ \frac{\pi \Pi(x)}{w(x)} \biggr]^2
      + \varphi(x) \cA \varphi(x)
    \Biggr)
  } \!
\end{equation}
with the SL operator
\begin{equation}
\label{SL_op}
\cA
\equiv
- \frac{1}{w(x)} \bigl[ \partial_x p(x) \partial_x + q(x) \bigr],
\end{equation} \vspace{-3mm}
\begin{equation}
\label{wpq_defs}
w(x)
\equiv
\frac{K(x)}{v(x)},
\quad
p(x)
\equiv
v(x)K(x),
\quad
q(x) = 0,
\end{equation}
where $\cA$ acts on the space $\cD$ of smooth \cite{Note:C2} functions $u(x)$ on $[-L/2, L/2]$ that satisfy $v(x)K(x) \partial_x u(x)\big|_{x = \pm L/2} = 0$, i.e., Neumann BCs when $v(x), K(x) > 0$.

We want to solve the SL problem
\begin{equation}
\label{SL_problem}
\cA u = \lambda u,
\quad
u \in \cD.
\end{equation}
It is regular if certain conditions are satisfied \cite{Note:Reg_vs_irreg}, and irregular otherwise.
This matters due to properties that hold for regular problems \cite{Zettl:2010}:
\begin{enumerate}[label = (\roman*), topsep=0.75ex, itemsep=0.65ex, partopsep=1ex, parsep=0.75ex, leftmargin=5.25ex]

\item
\label{Item:SL1}
The spectrum consists of eigenvalues $\lambda_{n}$ that are real and ordered so that $0 \leq \lambda_{0} < \lambda_{1} < \ldots$.

\item
\label{Item:SL2}
Each $\lambda_{n}$ has a unique square-integrable eigenfunction $u_{n}(x)$ with $n$ isolated zeros in $[-L/2, L/2]$.

\item
\label{Item:SL3}
The $u_{n}(x)$ form a complete orthonormal basis for square-integrable functions on $[-L/2, L/2]$ with weight $w(x)$, i.e., $\int_{-L/2}^{L/2} \dd x\, w(x) u_{n}(x) u_{n'}(x)
= \delta_{n,n'}$.

\end{enumerate}
Furthermore, assuming that the integral defining $1/v_{0}$ in Eq.~\eqref{yx_v0_def} converges, Weyl's law implies the following:
\begin{enumerate}[label = (\roman*), topsep=0.75ex, itemsep=0.65ex, partopsep=1ex, parsep=0.75ex, leftmargin=5.25ex]
\setcounter{enumi}{3}

\item
\label{Item:SL4}
$\lambda_{n}$ grows at most as $n^2$ for $n \gg 1$, i.e., $\lambda_{n} = O(n^2)$;

\end{enumerate}
cf.\ the SM \cite{SM} and \cite{AtkinsonMingarelli:1987, NiessenZettl:1992}.

Unless otherwise stated, our SL problem will be regular.
If irregular, the above properties are not guaranteed, while in certain cases they survive.
Note also that, since $q(x) = 0$ here, $u_{0}(x) = \textrm{const}$ solves Eq.~\eqref{SL_problem} with $\lambda_{0} = 0$.

The stated properties allow us to expand
\begin{equation}
\label{varphi_Pi_expansions}
\varphi(x)
= \sum_{n = 0}^{\infty} \varphi_{n} u_{n}(x),
\qquad
\Pi(x)
= \sum_{n = 0}^{\infty} \Pi_{n} w(x) u_{n}(x)
\end{equation}
with $\bigl[ \varphi_{n\ppr}, \Pi_{n'} \bigr] = \ii \delta_{n,n'}$ and thereby diagonalize $H_{\textrm{iTLL}}$ in Eq.~\eqref{H_iTLL}.
Indeed, the latter becomes
\begin{equation}
\label{H_iTLL_aa}
H_{\textrm{iTLL}}
= \frac{1}{2\pi} \bigl[ \pi \Pi_{0} \bigr]^2
  + \sum_{n = 1}^{\infty}
    E_{n} a^\dagger_{n} a\pdag_{n}
\end{equation}
with eigenenergies
\begin{equation}
\label{E_n}
E_{n} = \sqrt{\lambda_{n}},
\end{equation}
where we expressed $\varphi_{n} = \sqrt{\pi/2E_{n}}\left( a\pdag_{n} + a^\dagger_{n} \right)$ and $\Pi_{n} = -\ii \sqrt{E_{n}/2\pi}  \left( a\pdag_{n} - a^\dagger_{n} \right)$ for $n \geq 1$ in bosonic operators $a^\dagger_{n}$, $a\pdag_{n}$ satisfying $\bigl[ a\pdag_{n\ppr}, a^\dagger_{n'} \bigr] = \delta_{n,n'}$ \cite{Note:Zero_modes}.
\begin{figure*}[t]
\centering
\vspace{-5mm}
\includegraphics[width=\textwidth]{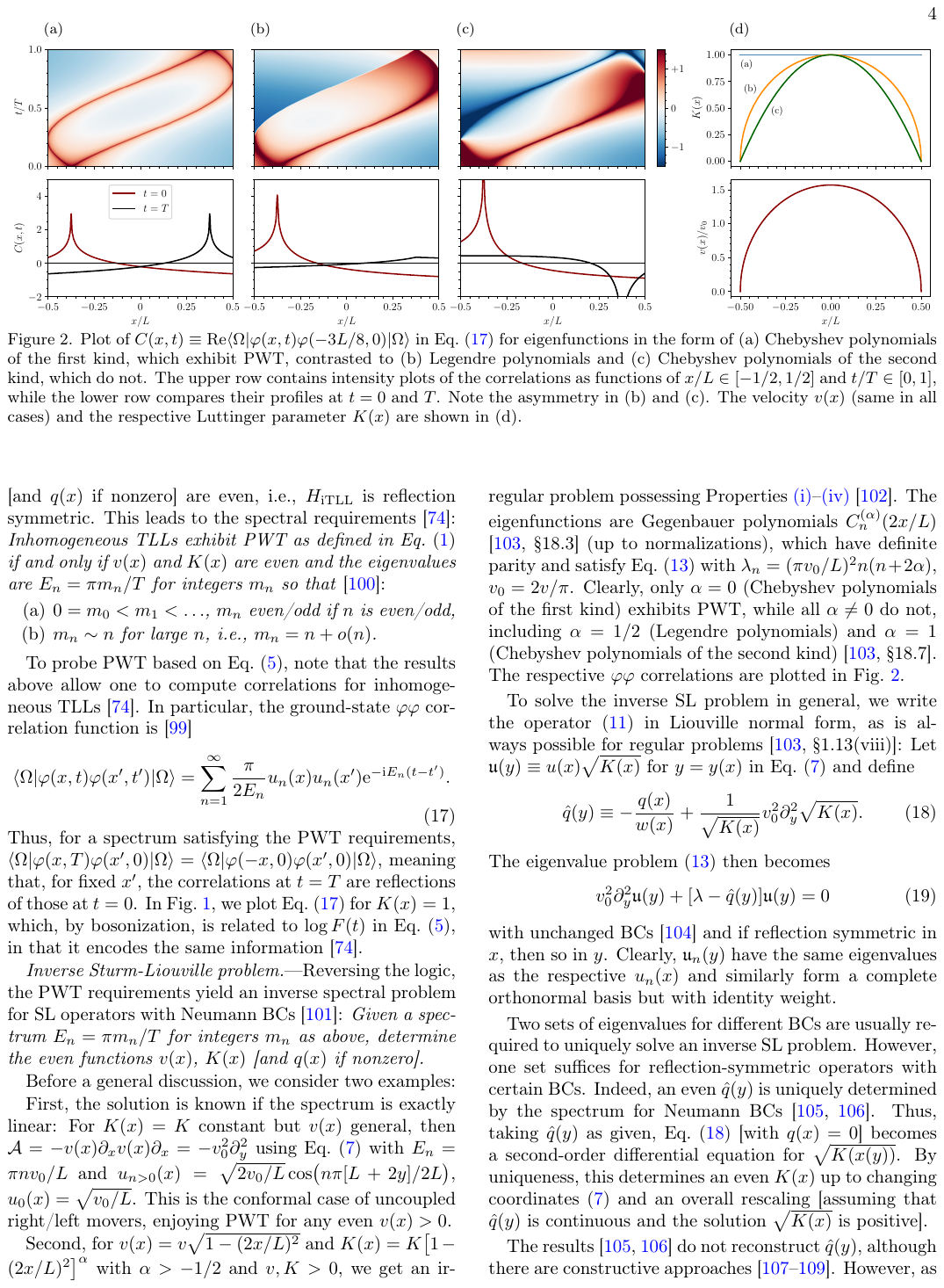}
\vspace{-5mm}
\caption{%
Plot of $C(x, t) \equiv \Re \langle\Omega| \varphi(x,t)\varphi(-3L/8,0) |\Omega\rangle$ given by Eq.~\eqref{varphivarphi} for eigenfunctions in the form of (a) Chebyshev polynomials of the first kind, which exhibit PWT, contrasted with (b) Legendre polynomials and (c) Chebyshev polynomials of the second kind, which do not.
The upper row contains intensity plots of the correlations as functions of $x/L \in [-1/2, 1/2]$ and $t/T \in [0,1]$, while the lower row compares their profiles at $t = 0$ and $T$.
Note the asymmetry in (b) and (c).
The velocity $v(x)$ (same in all cases) and the respective Luttinger parameter $K(x)$ are shown in (d).%
}%
\label{Fig:MCGS_fig2}
\vspace{-3mm}
\end{figure*}
The SM \cite{SM} also examines operators of the form \eqref{SL_op} with $q(x) \neq 0$, corresponding to an inhomogeneous (possibly imaginary) mass $M(x)$ by adding $v_{0}^2 M(x)^2 \varphi(x)^2$ inside the big parentheses in Eq.~\eqref{H_iTLL}.

The time evolution of any observable, such as $\varphi(x)$, contains terms like $a^{\dagger}_{n} \ee^{\ii E_{n} t} u_{n}(x)$ and $a\pdag_{n} \ee^{-\ii E_{n} t} u_{n}(x)$.
Thus, for Eq.~\eqref{PWT_def} to hold for all states, each $u_{n}(x)$ must have definite parity, i.e., be invariant under reflections $x \mapsto -x$ up to a sign, which must equal each phase factor $\ee^{\pm \ii E_{n} T}$ for a common shortest possible $T > 0$.
Moreover, the $u_{n}(x)$ have definite parity if and only if $v(x)$, $K(x)$ [and $q(x)$ if nonzero] are even, i.e., $H_{\textrm{iTLL}}$ is reflection symmetric.
This leads to a statement in terms of spectral requirements \cite{SM}:
\emph{Inhomogeneous TLLs exhibit PWT as defined in Eq.~\eqref{PWT_def} if and only if $v(x)$ and $K(x)$ are even and the eigenenergies are $E_{n} = \pi m_{n}/T$ for integers $m_{n}$ so that} \cite{Note:Prop_details}
\begin{enumerate}[label = \textnormal{(\alph*)}, topsep=0.75ex, itemsep=0.0ex, partopsep=1ex, parsep=0.75ex]

\item
\label{Item:mn_TLL_a}
\emph{$0 = m_{0} < m_{1} < \ldots$ with each $m_{n}$ even or odd if $n$ is even or odd,}

\item
\label{Item:mn_TLL_b}
\emph{$m_{n} \sim n$ for large $n$, i.e., $m_{n} = n + o(n)$.}

\end{enumerate}

To probe PWT based on Eq.~\eqref{Ft}, note that the results above allow one to compute correlations for inhomogeneous TLLs \cite{SM}.
In particular, the ground-state $\varphi\varphi$ correlation function is \cite{Note:Zero_modes}
\begin{equation}
\label{varphivarphi}
\langle\Omega| \varphi(x,t)\varphi(x',t') |\Omega\rangle
= \sum_{n = 1}^{\infty} \frac{\pi}{2 E_{n}}
  u_{n}(x) u_{n}(x') \ee^{-\ii E_{n}(t-t')}.
\end{equation}
Thus, for a spectrum satisfying the PWT requirements,
$\langle\Omega| \varphi(x,T)\varphi(x',0) |\Omega\rangle = \langle\Omega| \varphi(-x,0)\varphi(x',0) |\Omega\rangle$,
meaning that, for fixed $x'$, the correlations at $t = T$ are reflections of those at $t = 0$.
In Fig.~\ref{Fig:MCGS_fig1}, we plot Eq.~\eqref{varphivarphi} for $K(x) = 1$, which, by bosonization, is related to $\log F(t)$ in Eq.~\eqref{Ft}, in that it encodes the same information \cite{SM}.

\emph{Inverse Sturm-Liouville problem}---%
\csname phantomsection\endcsname%
\addcontentsline{toc}{section}{Inverse Sturm-Liouville problem}%
%
Reversing the logic, the spectral requirements for PWT yield an inverse problem for SL operators with Neumann BCs \cite{Note:GLM_eq}:
\emph{Given a spectrum $E_{n} = \pi m_{n}/T$ for integers $m_{n}$ as above, determine the even functions $v(x)$, $K(x)$ \textnormal{[}and $q(x)$ if nonzero\textnormal{]}.}

Before a general discussion, we consider two examples.

First, the solution is known if the spectrum is exactly linear:
For $K(x) = K$ constant but $v(x)$ general, then $\cA = - v(x) \partial_{x} v(x) \partial_{x} = -v_{0}^2 \partial_{y}^2$ using Eq.~\eqref{yx_v0_def} with $E_{n} = \pi n v_{0}/L$ and $u_{n > 0}(x) = \sqrt{2 v_{0}/L} \cos \bigl( n\pi [L + 2y]/2L \bigr)$, $u_{0}(x) = \sqrt{v_{0}/L}$.
This is the conformal case of uncoupled right and left movers, enjoying PWT for any even $v(x) > 0$.

Second, for $v(x) = v \sqrt{1 - (2x/L)^2}$ and $K(x) = K \bigl[ 1 - (2x/L)^2 \bigr]^\alpha$ with $\alpha > -1/2$ and $v, K > 0$, we get an irregular problem possessing Properties \ref{Item:SL1}--\ref{Item:SL4} \cite{Note:Irreg_BCs}.
The eigenfunctions are Gegenbauer polynomials $C^{(\alpha)}_{n}(2x/L)$ (up to normalizations), which have definite parity and satisfy Eq.~\eqref{SL_problem} with $\lambda_{n} = (\pi v_{0}/L)^2 n(n + 2\alpha)$, $v_{0} = 2v/\pi$; cf.\ Sec.~18.3 in \cite{NIST:DLMF}.
Clearly, only $\alpha = 0$ (Chebyshev polynomials of the first kind) exhibits PWT, while all $\alpha \neq 0$ do not, including $\alpha = 1/2$ (Legendre polynomials) and $\alpha = 1$ (Chebyshev polynomials of the second kind); cf.\ Sec.~18.7 in \cite{NIST:DLMF}.
The respective $\varphi\varphi$ correlations are plotted in Fig.~\ref{Fig:MCGS_fig2}.

To solve the inverse SL problem in general, we write the operator \eqref{SL_op} in Liouville normal form, as is always possible for regular problems; cf.\ Sec.~1.13(viii) in \cite{NIST:DLMF}:
Let $\mfu(y) \equiv u(x) \sqrt{K(x)}$ for $y = y(x)$ in Eq.~\eqref{yx_v0_def} and define
\begin{equation}
\label{qhat}
\hat{q}(y)
\equiv
- \frac{q(x)}{w(x)} + \frac{1}{\sqrt{K(x)}} v_{0}^2 
 \partial_{y}^2 \sqrt{K(x)}.
\end{equation}
The eigenvalue problem \eqref{SL_problem} then becomes
\begin{equation}
\label{SL_problem_mfu}
v_{0}^2 \partial_{y}^2 \mfu(y) + [\lambda - \hat{q}(y)] \mfu(y) = 0
\end{equation}
with unchanged BCs \cite{Note:LNF_BCs}, and if reflection symmetric in $x$, then so in $y$.
Clearly, $\mfu_{n}(y)$ have the same eigenvalues as the respective $u_{n}(x)$, and they similarly form a complete orthonormal basis but with identity weight.

Two sets of eigenvalues for different BCs are usually required to uniquely solve an inverse SL problem.
However, one set suffices for reflection-symmetric operators with certain BCs.
Indeed, an even $\hat{q}(y)$ is uniquely determined by the spectrum for Neumann BCs \cite{Borg:1946, Levinson:1949}.
Thus, taking $\hat{q}(y)$ as given, Eq.~\eqref{qhat} [with $q(x) = 0$] becomes a second-order differential equation for $\sqrt{K(x(y))}$.
By uniqueness, this determines an even $K(x)$ up to changing coordinates as in Eq.~\eqref{yx_v0_def} and an overall rescaling [assuming that $\hat{q}(y)$ is continuous and the solution $\sqrt{K(x)}$ is positive].

The results \cite{Borg:1946, Levinson:1949} do not reconstruct $\hat{q}(y)$, although there are constructive approaches \cite{Hald:1978constructive, Hochstadt:1975constructive, Gladwell:1986}.
However, as $K(x) = K$ gives an SL problem with $m_{n} = n$ and $\hat{q}(y) = 0$, a constant Luttinger parameter is the unique solution for the inverse problem given an exactly linear spectrum.
Beyond this case, one may imagine that the spectrum could deviate from the leading $n$ behavior as long as the requirements on the $m_{n}$ are satisfied.
By an exact Bohr-Sommerfeld/WKB quantization condition \cite{Dunham:1932, BenderEtAl:1977, Frasca:2007, BrackBhaduri:2003}, we show in the SM \cite{SM} that this depends on the regularity:
If the system is regular also after unfolding the interval $[-L/2, L/2]$ to the circle of twice the length \cite{Note:Unfolding_K(x)}, then PWT requires $\hat{q}(y) = 0$.
Thus, the conformal case of an exactly linear spectrum is the only possibility for such regular inhomogeneous TLLs to exhibit PWT.

\emph{Discussion}---%
\csname phantomsection\endcsname%
\addcontentsline{toc}{section}{Discussion}%
%
We investigated the collective transfer of quantum information in continuous communication channels, introducing the notion of perfect wave transfer (PWT).
Assuming conformal invariance, we showed that PWT is exhibited for one-particle excitations whenever the system is reflection symmetric.
More generally, PWT with respect to arbitrary states amounts to an inverse spectral problem, studied in detail for inhomogeneous TLL theory---a prototypical model of 1+1D bosons with coupled right and left movers.
Its solution shows that conformal invariance is essential for PWT if sufficient regularity is required.
This approach is similar to results for spin chains with nearest-neighbor hopping:
Their Hamiltonians can be represented as Jacobi matrices, yielding an inverse eigenvalue problem \cite{Kay:2010, ChristandlEtAl:2017} whose solution is unique \cite{Hochstadt:1974, Gladwell:1986} given one set of eigenvalues for matrices symmetric with respect to the antidiagonal (reflection symmetry) \cite{Hald:1976}.
Remarkably, however, the continuum problem is more constrained due to asymptotic requirements, which imply that the spectrum must be exactly linear for inhomogeneous bosons to exhibit PWT under strong enough regularity conditions.
This is not the case for discrete systems due to their intrinsic ultraviolet cutoff.
Such a cutoff is also relevant for applications of TLL theory \cite{TajikEtAl:2023}, and if present, there are no asymptotic requirements and more possibilities for PWT even for regular inhomogeneous TLLs.
Moreover, as shown in the SM \cite{SM}, PWT is also possible in inhomogeneous TLLs with fine-tuned mass terms, governed by a similar inverse spectral problem but with $q(x) \neq 0$.

By bosonization, our results for inhomogeneous TLLs cover fermionic theories with position-dependent interactions \cite{Moosavi:DBdG_iQLs:2023}.
The essential property for these systems to exhibit PWT as in Eq.~\eqref{PWT_def} was shown to be conformal invariance, assuming that the inhomogeneities are sufficiently regular, while it remains unclear what is possible if the regularity is relaxed.
Consequently, PWT for all (not only one-particle) states holds for particular CFTs, such as inhomogeneous CFTs of free fermions and compactified bosons.
It would be interesting to prove if PWT holds for any number of excitations in general CFTs and to extend to systems on the half or full infinite line.

Our Letter can be seen as a first investigation of PWT for continuous quantum systems, focusing on universal low-energy descriptions within CFT and TLL theory.
A second step would be to study PWT for specific gapless 1+1D models, including Lieb-Liniger \cite{LiebLiniger:1963}.
For those, our results apply at low energies, while full treatments require tailored solution methods, such as Bethe ansatz, but typically extended to inhomogeneous systems for PWT to be possible---a very difficult problem in itself.
Generic interaction or curvature effects may also spoil PWT beyond the low-energy regime.
Refinements may then be needed, either relaxing PWT to specific classes of observables or states, or reformulating it as \emph{almost PWT} with fidelity $1 - \varepsilon$ ($\varepsilon > 0$) for the transfer of information.
A concrete proposal for a general approach that builds on our present results would be to use nonlinear TLL theory \cite{ImambekovGlazman:2009, ImambekovSchmidtGlazman:2012} to study such questions.

\emph{Acknowledgments}---%
\csname phantomsection\endcsname%
\addcontentsline{toc}{section}{Acknowledgments}%
%
We are thankful to Pawel Caputa, Shinsei Ryu, and Imke Schneider for motivating discussions.
P.M.\ acknowledges financial support from the Wenner-Gren Foundations (Grant No.\ FT2022-0002).
M.C.\ acknowledges financial support from the European Research Council (Grant No.\ 818761), Villum Fonden via the QMATH Centre of Excellence (Grant No.\ 10059), and the Novo Nordisk Foundation (Grant No.\ NNF20OC0059939 ``Quantum for Life'') and thanks the NCCR SwissMAP of the Swiss National Science Foundation and the Section of Mathematics at the University of Geneva for their hospitality.
We also thank the organizers of the workshop ``Integrability in Condensed Matter Physics and Quantum Field Theory'' in 2023 at the SwissMAP Research Station in Les Diablerets, where this work was initiated.


\setlength{\bibsep}{-0.13ex}

%

\onecolumngrid
\clearpage

\begin{center}
{\Large Supplemental Material for:\vspace{0.2mm}\\
Perfect Wave Transfer in Continuous Quantum Systems}
\end{center}
\csname phantomsection\endcsname%
\addcontentsline{toc}{part}{Supplemental Material}%


This supplemental material has three parts.
Part~A gives details for two-point correlations in inhomogeneous 1+1D conformal field theory (CFT) with open boundary conditions (BCs).
Part~B describes how to diagonalize general inhomogeneous Tomonaga-Luttinger liquids (TLLs) using Sturm-Liouville (SL) theory, leading to the formulation of perfect wave transfer (PWT) as an inverse SL problem, including extension to ``massive'' TLLs and consequences inferred from Weyl's law and the Bohr-Sommerfeld/WKB quantization condition.
Part~C contains details on correlations in inhomogeneous TLLs.

\vspace{-4mm}
\section*{Part~A: Correlations in inhomogeneous 1+1D CFT with open boundaries}
\vspace{-3mm}

The general inhomogeneous CFT Hamiltonian is (setting $\hbar = 1$)
\begin{equation}
\label{H_iCFT}
H_{\textrm{iCFT}}
\equiv
\int_{-L/2}^{L/2} \dd x\, v(x) \bigl[ T_{+}(x) + T_{-}(x) \bigr],
\end{equation}
where $T_{\pm}(x)$ are the right- ($+$) and left- ($-$) moving components of the stress-energy tensor in light-cone coordinates, whose Fourier transforms yield two commuting copies of the Virasoro algebra; see, e.g., \cite{Moosavi:iCFT:2024}.
In position space, they satisfy the equal-time commutation relations
\begin{equation}
\bigl[ T_{\pm}(x), T_{\pm}(x') \bigr]
= \mp 2\ii \partial_{x} \delta(x-x') T_{\pm}(x')
  \pm \ii \delta(x-x') \partial_{x'}T_{\pm}(x')
  \pm \frac{c}{24\pi} \ii \partial_{x}^{3} \delta(x-x')
\end{equation}
and $\bigl[ T_{\pm}(x), T_{\mp}(x') \bigr] = 0$, where $c$ is the so-called central charge.
Note that $T_{\pm}$ should be understood to only depend on the ``right/left-moving coordinate'' $x^\mp$, i.e., $T_{\pm}(x^\mp)$; for now, we only view these as labels, distinguishing dependencies for right- or left-moving components.

Besides the $T_{\pm}(x)$ fields, an important class of operators are Virasoro primary fields $\Phi(x^-, x^+)$, defined by how they behave under conformal transformations.
The latter are given by orientation-preserving circle diffeomorphisms $f_+$ and $f_-$, under which primary fields transform as
\begin{equation}
\Phi(x^-, x^+)
\to
f'_+(x^-)^{\Delta_{\Phi}^+} f'_-(x^+)^{\Delta_{\Phi}^-} \Phi(f_+(x^-), f_-(x^+)),
\end{equation}
where ($\Delta_{\Phi}^+, \Delta_{\Phi}^-$) denote the field's conformal weights.
To impose open BCs at $x = \pm L/2$, as in the main text, we will assume that primary fields come in pairs $\Phi_{+}$, $\Phi_{-}$ with conformal weights satisfying $\Delta^+_{\Phi_{+}} = \Delta^-_{\Phi_{-}}$, $\Delta^-_{\Phi_{+}} = \Delta^+_{\Phi_{-}}$, since this allows one to couple excitations so that right movers are reflected to left movers when reaching $x = L/2$, and vice versa at $x = -L/2$.
For ease of notation, we will use $\Delta^+_{\Phi} \equiv \Delta^+_{\Phi_{+}}$, $\Delta^-_{\Phi} \equiv \Delta^-_{\Phi_{+}}$ for each such pair.

Before continuing, we mention two examples:
\begin{enumerate}[label = {\bf Example~\arabic*}:, ref = {\arabic*}, leftmargin=7.0em]

\item
\label{Item:Example1}
For the inhomogeneous free fermions in the main text, the components of the stress-energy tensor are
$T_{\pm}(x)
= (1/2) \bigl[
    \! \wick{ \psi^{\dagger}_{\pm}(x)
    (\mp \ii \partial_x ) \psi\pdag_{\pm}(x) } + \hc
  \bigr]
  - \pi \big/ 12L^2$,
the central charge is $c = 1$, and the primary fields of the theory are the fermionic fields $\psi_{\pm}(x)$ themselves, whose conformal weights are
$\Delta^+_{\psi_{+}} = \Delta^-_{\psi_{-}} = 1/2$, $\Delta^-_{\psi_{+}} = \Delta^+_{\psi_{-}} = 0$.

\item
\label{Item:Example2}
For an inhomogeneous TLL with constant Luttinger parameter $K(x) = K$, one has
$T_{\pm}(x)
= ({1}/{4\pi K})
  \! \wick{ \bigl[ \pi \Pi(x) \mp K \partial_x \varphi(x) \bigr]^2}
  - \pi \big/ 12L^2$,
the central charge is $c = 1$, and the primary fields include vertex operators constructed as exponentials of the bosonic fields.
Applied to the Luttinger model \cite{Tomonaga:1950, Luttinger:1963, MattisLieb:1965}, these vertex operators are (renormalized) fermionic fields $\psi_{\pm}$ with conformal weights $\Delta^+_{\psi_{+}} = \Delta^-_{\psi_{-}} = (1+K)^2/8K$, $\Delta^-_{\psi_{+}} = \Delta^+_{\psi_{-}} = (1-K)^2/8K$.
(Free fermions are recovered by setting $K = 1$.)

\end{enumerate}

As in the main text, open BCs can be incorporated by standard unfolding methods; cf.\ \cite{EggertAffleck:1992, FabrizioGogolin:1995, Giamarchi:2003, Cazalilla:2004, GawedzkiKozlowski:2020, ChuaEtAl:2020}:
Couple right- and left-moving excitations by identifying
\begin{equation}
\label{unfolding_rels}
T_{-}(x) = T_{+}(L-x),
\qquad
\Phi_{-}(x^-, x^+) = \Phi_{+}(L-x^+, L-x^-)
\end{equation}
for all pairs $\Phi_{+}, \Phi_{-}$ of Virasoro primary fields.
It follows that we can equivalently consider a chiral inhomogeneous CFT of only right movers on the interval $[-L/2, 3L/2]$ of length $2L$ with velocity
\begin{equation}
\uv(x)
\equiv
\begin{cases}
  v(x) & \text{for } x \in [-L/2, L/2], \\
  v(L-x) & \text{for } x \in [L/2, 3L/2],
\end{cases}
\end{equation}
and with every quantity satisfying periodic BCs.
Note that the Hamiltonian \eqref{H_iCFT} can then be written as
\begin{equation}
\label{H_iCFT_2L}
H_{\textrm{iCFT}}
\equiv
\int_{-L/2}^{3L/2} \dd x\, \uv(x) T_{+}(x).
\end{equation}

For later reference, we define
\begin{equation}
\uf(x) 
\equiv \int_{0}^{x} \dd s\, \frac{\uv_{0}}{\uv(s)},
\qquad
\frac{1}{\uv_{0}}
\equiv \frac{1}{2L} \int_{-L/2}^{3L/2} \frac{\dd s}{\uv(s)},
\qquad
\ux_{t}^{\pm}(x)
\equiv \uf^{-1} \bigl( \uf(x) \pm \uv_{0}t \bigr).
\end{equation}
Naturally, $\uf$ lies in the universal cover $\wDiff_{+}(S^1_{2L})$ of the group of orientation-preserving diffeomorphisms of the circle $S^1_{2L}$ with circumference $2L$. 
Given any $v(x)$, one can show the following properties:
\begin{equation}
\label{vbar_fbar_props}
\uv(L-x) = \uv(x),
\qquad
\uf(L-x) = \uf(L) - \uf(x),
\qquad
\uf^{-1} \bigl( \uf(L)-y \bigr) = L - \uf^{-1}(y),
\qquad
\ux_{t}^{\pm}(L - x) = L - \ux_{t}^{\mp}(x).
\end{equation}
Lastly, we introduce the undeformed Hamiltonian
\begin{equation}
H_{0} = \int_{-L/2}^{3L/2} \dd x\, \uv_{0} T_{+}(x),
\end{equation}
whose ground state we denote by $|\Omega_{0}\rangle$.
Similarly, we let $|\Omega\rangle$ denote the ground state of $H_{\textrm{iCFT}}$.
Following \cite{GLM:2018, Moosavi:iCFT:2024}, the latter are related to their undeformed counterparts by
\begin{equation}
\label{H_iCFT_transf_rules_Omega}
U_{+}(\uf) H_{\textrm{iCFT}} U_{+}(\uf)^{-1}
= H_{0} + \textrm{const},
\qquad
|\Omega\rangle
= U_{+}(\uf)^{-1} |\Omega_{0}\rangle,
\end{equation}
where $U_{+}(\uf)$ is a projective unitary representation of $\uf$ on the Hilbert space of the CFT.

Similar to the main text, we define
\begin{equation}
|\Phi_{j}\rangle
\equiv
\int_{-L/2}^{L/2} \dd x\, \left( \ee^{-\ii k x} \xi^{+}_{j}(x) \Phi_{+}^\dagger(x,x) + \ee^{\ii k x} \xi^{-}_{j}(x) 
 \Phi_{-}^\dagger(x,x) \right) |\Omega\rangle
\qquad (j = 1,2)
\end{equation}
for smooth enough functions $\xi^{\pm}_{j}(x)$, interpreted as ``waves'' of right- or left-moving excitations in each state, and some $k \in \mathbb{R}$ in order to accommodate for $\kF$ in Eq.~\eqref{psi_def}.
(The latter will have no effect on PWT, and one might as well set $k = 0$.)
As in the main text, this is a restricted class of states which only allow for one-particle excitations, while we eased other restrictions in the sense that the excitations are not necessarily fermionic and the distributions of right and left movers in the ``waves'' can be different.

To probe PWT for one-particle excitations, we seek to compute
\begin{multline}
\langle \Phi_{j}| \Phi_{j}\rangle
= \int_{-L/2}^{L/2} \dd x_1\, \int_{-L/2}^{L/2} \dd x_2\,
\langle\Omega|
  \left( \ee^{\ii k x_2} \overline{\xi^{+}_{j}(x_2)} \Phi_{+}\pdag(x_2,x_2) + \ee^{-\ii k x_2} \overline{\xi^{-}_{j}(x_2)} \Phi_{-}\pdag(x_2,x_2) \right) \\
\times
  \left( \ee^{-\ii k x_1} \xi^{+}_{j}(x_1) \Phi_{+}^\dagger(x_1,x_1) + \ee^{\ii k x_1} \xi^{-}_{j}(x_1) \Phi_{-}^\dagger(x_1,x_1) \right)
|\Omega\rangle
\end{multline}
and
\begin{multline}
F(t)
\equiv
\langle \Phi_{2}| \ee^{-\ii H_{\textrm{iCFT}}t} |\Phi_{1}\rangle
= \int_{-L/2}^{L/2} \dd x_1\, \int_{-L/2}^{L/2} \dd x_2\,
\langle\Omega|
  \left( \ee^{\ii k x_2} \overline{\xi^{+}_{2}(x_2)} \Phi_{+}\pdag(x_2,x_2) + \ee^{-\ii k x_2} \overline{\xi^{-}_{2}(x_2)} \Phi_{-}\pdag(x_2,x_2) \right)
  \ee^{-\ii H_{\textrm{iCFT}}t} \\
\times
  \left( \ee^{-\ii k x_1} \xi^{+}_{1}(x_1) \Phi_{+}^\dagger(x_1,x_1) + \ee^{\ii k x_1} \xi^{-}_{1}(x_1) \Phi_{-}^\dagger(x_1,x_1) \right)
|\Omega\rangle,
\end{multline}
in order to eventually compare $|F(t)|$ with $|\langle \Phi_{1}| \Phi_{1}\rangle|$ for specular waves, $\xi^{\pm}_{2}(x) = \xi^{\mp}_{1}(-x)$.
Using the unfolding relations in Eq.~\eqref{unfolding_rels}, we obtain
\begin{multline}
\langle \Phi_{j}| \Phi_{j}\rangle
= \int_{-L/2}^{L/2} \dd x_1\, \int_{-L/2}^{L/2} \dd x_2\,
\langle\Omega|
  \left( \ee^{\ii k x_2} \overline{\xi^{+}_{j}(x_2)} 
 \Phi_{+}\pdag(x_2,x_2) + \ee^{-\ii k x_2} \overline{\xi^{-}_{j}(x_2)} \Phi_{+}\pdag(L-x_2,L-x_2) \right) \\
\times
  \left( \ee^{-\ii k x_1} \xi^{+}_{j}(x_1) \Phi_{+}^\dagger(x_1,x_1) + \ee^{\ii k x_1} \xi^{-}_{j}(x_1) \Phi_{+}^\dagger(L-x_1,L-x_1) \right)
|\Omega\rangle
\end{multline}
and
\begin{multline}
F(t)
= \ee^{-\ii Ct} \int_{-L/2}^{L/2} \dd x_1\, \int_{-L/2}^{L/2} \dd x_2\,
\langle\Omega|
  \ee^{\ii H_{\textrm{iCFT}}t}
  \left( \ee^{\ii k x_2} \overline{\xi^{+}_{2}(x_2)} \Phi_{+}\pdag(x_2,x_2) + \ee^{-\ii k x_2} \overline{\xi^{-}_{2}(x_2)} \Phi_{+}\pdag(L-x_2,L-x_2) \right)
  \ee^{-\ii H_{\textrm{iCFT}}t} \\
\times
  \left( \ee^{-\ii k x_1} \xi^{+}_{1}(x_1) \Phi_{+}^\dagger(x_1,x_1) + \ee^{\ii k x_1} \xi^{-}_{1}(x_1) \Phi_{+}^\dagger(L-x_1,L-x_1) \right)
|\Omega\rangle,
\end{multline}
where we used that $H_{\textrm{iCFT}} |\Omega\rangle = C |\Omega\rangle$ for some real constant $C$ that encodes the Casimir energy.
(This only gives rise to an overall phase that one could safely ignore.)
Clearly, as stated in the main text, the above amounts to evaluating and comparing two-point correlations for primary fields, specifically those of right movers in a chiral CFT on $[-L/2, 3L/2]$.

First, as a preliminary step, for primary fields in a system of length $2L$ with periodic BCs, conformal invariance implies that the (undeformed) two-point correlations are of the form
\begin{equation}
\langle\Omega_{0}| \Phi\pdag_{+}(x_2^-, x_2^+) \Phi^\dagger_{+}(x_1^-, x_1^+) |\Omega_{0}\rangle
= \frac{1}{2\pi}
  G^+_{\Phi\Phi}(x_2^- - x_1^-)^{2\Delta^+_{\Phi}}
  G^-_{\Phi\Phi}(x_2^+ - x_1^+)^{2\Delta^-_{\Phi}},
\end{equation}
where $G^\pm_{\Phi\Phi}(x)$ are $2L$-periodic functions.
For example, as can be seen from, e.g., \cite{DelftSchoeller:1998, LaMo1:2015}, the two-point correlations for right-moving free fermions (cf.\ Example~\ref{Item:Example1} above) are
\begin{equation}
\langle\Omega_{0}| \psi\pdag_{+}(x_2) \psi^\dagger_{+}(x_1) |\Omega_{0}\rangle
= \frac{1}{2\pi} G^+_{\textrm{FF}}(x_2 - x_1)
\end{equation}
with
\begin{equation}
\label{G_FF}
G^+_{\textrm{FF}}(x)
\equiv
\frac{
  \ii\pi \exp \bigl( -\ii \frac{\pi}{2L} x \bigr)
}{
  2L \sin\bigl( \frac{\pi}{2L} [x + \ii0^+] \bigr)
},
\end{equation}
where we used that $\Delta^+_{\psi} = 1/2$, $\Delta^-_{\psi} = 0$.
Similarly, for the Luttinger model (cf.\ Example~\ref{Item:Example2} above), the two-point correlations for the (renormalized) right-moving fermionic fields are
\begin{equation}
\langle\Omega_{0}| \psi\pdag_{+}(x_2^-, x_2^+) \psi^\dagger_{+}(x_1^-, x_1^+) |\Omega_{0}\rangle
= \frac{1}{2\pi}
  G^+_{\textrm{FF}}(x_2^- - x_1^-)^{(1+K)^2/4K}
  G^+_{\textrm{FF}}(x_1^+ - x_2^+)^{(1-K)^2/4K},
\end{equation}
where we used that
\begin{equation}
G^-_{\textrm{FF}}(x)
\equiv
\frac{
  -\ii\pi \exp \bigl( +\ii \frac{\pi}{2L} x \bigr)
}{
  2L \sin\bigl( \frac{\pi}{2L} [x - \ii0^+] \bigr)
}
= \frac{
    \ii\pi \exp \bigl( - \ii \frac{\pi}{2L} [-x] \bigr)
  }{
    2L
    \sin\bigl( \frac{\pi}{2L} [- x + \ii0^+] \bigr)
  }
= G^+_{\textrm{FF}}(-x).
\end{equation}

Second, by results spelled out in Proposition 3.1 in \cite{Moosavi:iCFT:2024} for inhomogeneous CFT, the time dependence in $F(t)$ means that those correlations are evaluated at spacetime points given by $\ux^{\pm}_t(x)$ in Eq.~\eqref{vbar_fbar_props}.
In addition, the action of $U_{+}(\uf)$ from the expression \eqref{H_iCFT_transf_rules_Omega} for $|\Omega\rangle$ also leads to a conformal transformation of the fields.
It follows from this and the properties in Eq.~\eqref{vbar_fbar_props} that
\begin{multline}
\label{PhiPhi_final}
\langle \Phi_{j}| \Phi_{j}\rangle
= \frac{1}{2\pi} \int_{-L/2}^{L/2} \dd x_1\, \int_{-L/2}^{L/2} \dd x_2\,
\uf'(x_2)^{\Delta^+_{\Phi}}\uf'(x_1)^{\Delta^+_{\Phi}}
\uf'(x_2)^{\Delta^-_{\Phi}}\uf'(x_1)^{\Delta^-_{\Phi}} \\
\begin{aligned}
\times
\biggl(
&
\ee^{\ii k (x_2-x_1)} \overline{\xi^{+}_{j}(x_2)} \xi^{+}_{j}(x_1)
G^+_{\Phi\Phi}\bigl(\uf(x_2) - \uf(x_1)\bigr)^{2\Delta^+_{\Phi}}
G^-_{\Phi\Phi}\bigl(\uf(x_2) - \uf(x_1)\bigr)^{2\Delta^-_{\Phi}} \\
& +
\ee^{\ii k (x_2+x_1)} \overline{\xi^{+}_{j}(x_2)} \xi^{-}_{j}(x_1)
G^+_{\Phi\Phi}\bigl(\uf(x_2) + \uf(x_1) - \uf(L)\bigr)^{2\Delta^+_{\Phi}}
G^-_{\Phi\Phi}\bigl(\uf(x_2) + \uf(x_1) - \uf(L)\bigr)^{2\Delta^-_{\Phi}} \\
& +
\ee^{-\ii k (x_2+x_1)} \overline{\xi^{-}_{j}(x_2)} \xi^{+}_{j}(x_1)
G^+_{\Phi\Phi}\bigl( - \uf(x_2) - \uf(x_1) + \uf(L)\bigr)^{2\Delta^+_{\Phi}}
G^-_{\Phi\Phi}\bigl( - \uf(x_2) - \uf(x_1) + \uf(L)\bigr)^{2\Delta^-_{\Phi}} \\
& +
\ee^{-\ii k (x_2-x_1)} \overline{\xi^{-}_{j}(x_2)} \xi^{-}_{j}(x_1)
G^+_{\Phi\Phi}\bigl( - \uf(x_2) + \uf(x_1)\bigr)^{2\Delta^+_{\Phi}}
G^-_{\Phi\Phi}\bigl( - \uf(x_2) + \uf(x_1)\bigr)^{2\Delta^-_{\Phi}}
\biggr)
\end{aligned}
\end{multline}
and
\begin{multline}
\label{Ft_final}
\hspace{-2.5mm}F(t)
= \ee^{-\ii Ct} \frac{1}{2\pi} \int_{-L/2}^{L/2} \dd x_1\, \int_{-L/2}^{L/2} \dd x_2\,
\uf'(x_2)^{\Delta^+_{\Phi}}\uf'(x_1)^{\Delta^+_{\Phi}}
\uf'(x_2)^{\Delta^-_{\Phi}}\uf'(x_1)^{\Delta^-_{\Phi}} \\
\begin{aligned}
\times
\biggl(
&
\ee^{\ii k (x_2-x_1)} \overline{\xi^{+}_{2}(x_2)} \xi^{+}_{1}(x_1) 
G^+_{\Phi\Phi}\bigl(\uf(x_2) - \uf(x_1) - \uv_{0}t\bigr)^{2\Delta^+_{\Phi}}
G^-_{\Phi\Phi}\bigl(\uf(x_2) - \uf(x_1) + \uv_{0}t\bigr)^{2\Delta^-_{\Phi}} \\
& +
\ee^{\ii k (x_2+x_1)} \overline{\xi^{+}_{2}(x_2)} \xi^{-}_{1}(x_1) 
G^+_{\Phi\Phi}\bigl(\uf(x_2) + \uf(x_1) - \uf(L) - \uv_{0}t\bigr)^{2\Delta^+_{\Phi}}
G^-_{\Phi\Phi}\bigl(\uf(x_2) + \uf(x_1) - \uf(L) + \uv_{0}t\bigr)^{2\Delta^-_{\Phi}} \\
& +
\ee^{-\ii k (x_2+x_1)} \overline{\xi^{-}_{2}(x_2)} \xi^{+}_{1}(x_1) 
G^+_{\Phi\Phi}\bigl( - \uf(x_2) - \uf(x_1) + \uf(L) - \uv_{0}t\bigr)^{2\Delta^+_{\Phi}}
G^-_{\Phi\Phi}\bigl( - \uf(x_2) - \uf(x_1) + \uf(L) + \uv_{0}t\bigr)^{2\Delta^-_{\Phi}} \\
& +
\ee^{-\ii k (x_2-x_1)} \overline{\xi^{-}_{2}(x_2)} \xi^{-}_{1}(x_1) 
G^+_{\Phi\Phi}\bigl( - \uf(x_2) + \uf(x_1) - \uv_{0}t\bigr)^{2\Delta^+_{\Phi}}
G^-_{\Phi\Phi}\bigl( - \uf(x_2) + \uf(x_1) + \uv_{0}t\bigr)^{2\Delta^-_{\Phi}}
\biggr),
\end{aligned}
\end{multline}
where we used that
$\uf\bigl(\ux_{t}^{\pm}(x) \bigr)
= \uf(x) \pm \uv_{0}t$
and
$\uf\bigl(\ux_{t}^{\pm}(L - x) \bigr)
= \uf(L) - \uf\bigl(\ux_{t}^{\mp}(x)\bigr)
= \uf(L) - \uf(x) \pm \uv_{0}t$.

Finally, for specular waves, $\xi^{\pm}_{2}(x) = \xi^{\mp}_{1}(-x)$, we obtain
\begin{multline}
\label{Ft_final_specular}
F(t)
= \ee^{-\ii Ct} \frac{1}{2\pi} \int_{-L/2}^{L/2} \dd x_1\, \int_{-L/2}^{L/2} \dd x_2\,
\uf'(-x_2)^{\Delta^+_{\Phi}}\uf'(x_1)^{\Delta^+_{\Phi}}
\uf'(-x_2)^{\Delta^-_{\Phi}}\uf'(x_1)^{\Delta^-_{\Phi}} \\
\begin{aligned}
\times
\biggl(
&
\ee^{\ii k (x_2-x_1)} \overline{\xi^{+}_{1}(x_2)} \xi^{+}_{1}(x_1)
G^+_{\Phi\Phi}\bigl( - \uf(-x_2) - \uf(x_1) + \uf(L) - \uv_{0}t\bigr)^{2\Delta^+_{\Phi}}
G^-_{\Phi\Phi}\bigl( - \uf(-x_2) - \uf(x_1) + \uf(L) + \uv_{0}t\bigr)^{2\Delta^-_{\Phi}} \\
& +
\ee^{\ii k (x_2+x_1)} \overline{\xi^{+}_{1}(x_2)} \xi^{-}_{1}(x_1)
G^+_{\Phi\Phi}\bigl( - \uf(-x_2) + \uf(x_1) - \uv_{0}t\bigr)^{2\Delta^+_{\Phi}}
G^-_{\Phi\Phi}\bigl( - \uf(-x_2) + \uf(x_1) + \uv_{0}t\bigr)^{2\Delta^-_{\Phi}} \\
& +
\ee^{-\ii k (x_2+x_1)} \overline{\xi^{-}_{1}(x_2)} \xi^{+}_{1}(x_1)
G^+_{\Phi\Phi}\bigl(\uf(-x_2) - \uf(x_1) - \uv_{0}t\bigr)^{2\Delta^+_{\Phi}}
G^-_{\Phi\Phi}\bigl(\uf(-x_2) - \uf(x_1) + \uv_{0}t\bigr)^{2\Delta^-_{\Phi}} \\
& +
\ee^{-\ii k (x_2-x_1)}  \overline{\xi^{-}_{1}(x_2)} \xi^{-}_{1}(x_1)
G^+_{\Phi\Phi}\bigl(\uf(-x_2) + \uf(x_1) - \uf(L) - \uv_{0}t\bigr)^{2\Delta^+_{\Phi}}
G^-_{\Phi\Phi}\bigl(\uf(-x_2) + \uf(x_1) - \uf(L) + \uv_{0}t\bigr)^{2\Delta^-_{\Phi}}
\biggr),
\end{aligned}
\end{multline}
after a change of variable and rearranging.

By comparing the terms appearing in Eqs.~\eqref{Ft_final_specular} and~\eqref{PhiPhi_final}, it follows that $|F(t)|$ for certain $t$ equals $|\langle \Phi_{1}| \Phi_{1}\rangle|$ for all $\xi^{\pm}_{1}(x)$ if and only if $\uv(x)$ is even, meaning $\uf(x)$ must be odd.
This in turn implies that $\uf(L) = L$ and $\uv_{0} = v_{0}$ given by Eq.~\eqref{yx_v0_def}, which means that equality holds for all $t = (2m+1)L/v_{0}$ due the $2L$ periodicity of $G^{\pm}_{\Phi\Phi}(x)$.
[Moreover, for $\xi^{\pm}_{2}(x) = \xi^{\mp}_{1}(-x)$, one has $|\langle \Phi_{1}| \Phi_{1}\rangle| = |\langle \Phi_{2}| \Phi_{2}\rangle|$ when $v(x)$ is even.]
Since $\uv(x) = v(x)$ for all $x \in [-L/2, L/2]$, this proves the claim that one-particle PWT is exhibited for all even inhomogeneous CFTs.

As a last remark, it is possible to unfold in different ways to incorporate the statistics of the primary fields.
E.g., so that, for the system of length $2L$, one gets antiperiodic BCs for fermions and twisted BCs for general anyonic fields.
However, since we are interested in recombinations of waves traveling in different directions, we found it necessary to identify right and left movers so as to always have periodic BCs.

\vspace{-4mm}
\section*{Part~B: Spectral properties of inhomogeneous TLLs}
\vspace{-3mm}

In this part, we first show that reflection symmetry is a necessary condition for PWT in any system; here, under the assumption that the Hamiltonian has pure point spectrum.
Second, we give the details for how to use SL theory to diagonalize the inhomogeneous TLL Hamiltonian \eqref{H_iTLL} and how to formulate the PWT requirements on the spectrum.
Third and fourth, respectively, we review certain details for Weyl's law and the Bohr-Sommerfeld/WKB quantization condition applied to SL operators, and, fifth, extend the previous discussion to TLL Hamiltonians with a mass term.
Sixth, we comment on the Liouville normal form used to state the results for the inverse SL problem in the main text.

\vspace{-3mm}
\subsection*{Necessity of reflection symmetry}
\vspace{-3mm}

Before considering TLL theory, we note that all models studied in the main text have pure point spectrum [as long as the integral defining $1/v_{0}$ in Eq.~\eqref{yx_v0_def} converges].
Thus, we first consider a general Hamiltonian $H$ with pure point spectrum and show that for such a system to exhibit PWT or PST, it is necessary that $H$ commutes with the parity operator $\cP$, which maps $x \mapsto -x$.
In other words, we will show that reflection symmetry is a necessary condition.

To prove the statement, let $E_{n}$ denote the eigenenergies of $H$ with the corresponding energy eigenstates $|n\rangle$.
On general grounds, the latter form a complete orthonormal basis for the Hilbert space $\cF$ of the theory.
Let $|\Phi_{1}\rangle$ be an arbitrary state in $\cF$ and let $|\Phi_{2}\rangle \equiv \cP |\Phi_{1}\rangle$.
(We recall that $\cP$ is a unitary operator satisfying $\cP^2 = 1$, meaning that $\cP^{-1} = \cP^\dagger = \cP$.)
As usual, define $F(t) \equiv \langle\Phi_{2}| \ee^{-\ii H t} |\Phi_{1}\rangle =  \langle\Phi_{1}| \cP \ee^{-\ii H t} |\Phi_{1}\rangle$.
Full PWT, as defined in Eq.~\eqref{PWT_def}, requires that there exists a common shortest possible time $T > 0$ such that $|F(T)| = |\langle\Phi_{1}|\Phi_{1}\rangle|$ for all states $|\Phi_{1}\rangle$.
In particular, this must hold for any energy eigenstate $|n\rangle$, implying that $|\langle n|\cP|n\rangle| = 1$.
If $|n\rangle$ is not an eigenstate of $\cP$, then it is straightforward to show that this equality cannot hold.
Consequently, all energy eigenstates must also be eigenstates of $\cP$, which implies that $[\cP, H] = 0$.

\vspace{-3mm}
\subsection*{Inhomogeneous TLLs and SL theory}
\vspace{-3mm}

We now turn to inhomogeneous TLL theory studied in the main text.
Inserting the expansions in Eq.~\eqref{varphi_Pi_expansions} into the Hamiltonian \eqref{H_iTLL_cA} and using Eq.~\eqref{SL_problem} yield
\begin{equation}
\label{H_iTLL_Fourier}
H_{\textrm{iTLL}}
= \frac{1}{2\pi} [\pi \Pi_{0}]^2
  + \frac{1}{2\pi} \sum_{n = 1}^{\infty} \wick{ \Bigl( [\pi \Pi_{n}]^2 + \lambda_{n} \varphi_{n}^2 \Bigr) }.
\end{equation}
Here, $\varphi_{0}$ is absent (when $E_{0} = 0$), no different from the standard case with constant $K(x)$ [and $v(x)$].
For $n \geq 1$, let
\begin{equation}
\label{varphi_Pi_a}
\varphi_{n}
= \sqrt{\frac{\pi}{2}} \frac{1}{\lambda_{n}^{1/4}} \left( a\pdag_{n} + a^\dagger_{n} \right),
\qquad
\Pi_{n}
= -\ii \frac{\lambda_{n}^{1/4}}{\sqrt{2\pi}} \left( a\pdag_{n} - a^\dagger_{n} \right)
\end{equation}
in terms of bosonic creation and annihilation operators $a^\dagger_{n}$ and $a\pdag_{n}$ satisfying $\bigl[ a\pdag_{n\ppr}, a^\dagger_{n'} \bigr] = \delta_{n,n'}$ and $a_{n} |\Omega\rangle = 0$, where $|\Omega\rangle$ denotes the ground state of $H_{\textrm{iTLL}}$; cf.\ Sec.~3.2 in \cite{GMS:2022}.
Inserting Eq.~\eqref{varphi_Pi_a} into Eq.~\eqref{H_iTLL_Fourier} yields Eqs.~\eqref{H_iTLL_aa}--\eqref{E_n}.

Using the above, one can compute the time evolution of any observable in the algebra of operators generated by the fields $\varphi(x)$ and $\Pi(x)$.
For instance, for $\varphi(x, t) \equiv \ee^{\ii H_{\textrm{iTLL}} t} \varphi(x) \ee^{-\ii H_{\textrm{iTLL}} t}$ and $\Pi(x, t)$ defined similarly, we obtain
\begin{equation}
\label{varphi_Pi_expansions_time}
\varphi(x, t)
= \sum_{n = 1}^{\infty} \sqrt{\frac{\pi}{2E_{n}}}
  \left( a\pdag_{n} \ee^{-\ii E_{n}t} + a^\dagger_{n} \ee^{\ii E_{n}t} \right)
  u_{n}(x),
\qquad
\Pi(x, t)
= \sum_{n = 1}^{\infty} (-\ii) \sqrt{\frac{E_{n}}{2\pi}}
  \left( a\pdag_{n} \ee^{-\ii E_{n}t} - a^\dagger_{n} \ee^{\ii E_{n}t} \right)
  w(x) u_{n}(x),
\end{equation}
where we omitted the zero modes (they must be treated separately when $E_{0} = 0$, but in that case they will have no consequence for our discussion of PWT) and used $\ee^{\ii H_{\textrm{iTLL}} t} a\pdag_{n} \ee^{-\ii H_{\textrm{iTLL}} t} = a\pdag_{n} \ee^{-\ii E_{n}t}$ and $\ee^{\ii H_{\textrm{iTLL}} t} a^\dagger_{n} \ee^{-\ii H_{\textrm{iTLL}} t} = a^\dagger_{n} \ee^{\ii E_{n}t}$.
Generally, any observable $\cO(x, t)$ in this algebra will contain contributions of the form $a^{\dagger}_{n} \ee^{\ii E_{n} t} u_{n}(x)$ and $a\pdag_{n} \ee^{-\ii E_{n} t} u_{n}(x)$.
Now, in order for PWT to be exhibited, Eq.~\eqref{PWT_def} must hold for any state in the Hilbert space, which means that it must hold without taking expectation values.
I.e., $\cO(x, 0)$ must equal $\cO(-x, T)$ at some later time $T > 0$ as an operator identity.
Taking $\cO = \varphi$ and using the linear independence of the generators $a^\dagger_{n}$ and $a\pdag_{n}$, Eq.~\eqref{varphi_Pi_expansions_time} implies that
\begin{equation}
\label{unx_unmx_T}
u_{n}(x) = u_{n}(-x) \ee^{\pm \ii E_{n} T}
\quad \forall n = 0, 1, \ldots
\end{equation}
for a common shortest possible $T > 0$, where we included the trivial case $n = 0$ [recall that $u_{0}(x) = \textrm{const}$ and $E_{0} = 0$ when $q(x) = 0$ in Eq.~\eqref{SL_op}].
Since $\cP u_{n}(x) = u_{n}(-x)$, Eq.~\eqref{unx_unmx_T} implies that each $u_{n}(x)$ must be an eigenfunction of the parity operator $\cP$, whose eigenvalues are $+1$ or $-1$ (since $\cP^2 = 1$).
It follows that all eigenfunctions must have definite parity, and the only possibility for PWT to hold is that the corresponding eigenenergies satisfy $E_{n} T = \pi m_{n}$ for integers $m_{n}$, with $m_{n}$ even (odd) if $u_{n}(x)$ has even (odd) parity.

Finally, it is straightforward to show the following statement:
The eigenfunctions $u_{n}(x)$ of a regular SL operator $\cA$ have definite parity if and only if $[\cA, \cP] = 0$.
For our operator in Eqs.~\eqref{SL_op}--\eqref{wpq_defs}, this means that both $v(x)$ and $K(x)$ must be even functions (using that they must be positive to exclude the possibility that both are instead odd).

Given the above, together with Properties~\ref{Item:SL1}--\ref{Item:SL4} for regular SL operators stated in the main text, we have derived our statement in terms of spectral requirements for an inhomogeneous TLL to exhibit PWT.

Moreover, as stated in the main text, reversing the logic, i.e., assuming we are given a spectrum that satisfies the spectral requirements for PWT, we obtain our formulation of PWT for inhomogeneous TLLs as an inverse problem.

\vspace{-3mm}
\subsection*{Weyl's law for SL operators}
\vspace{-3mm}

Weyl's law applied to SL operators of the general form \eqref{SL_op} states that the number $N(\lambda)$ of eigenvalues below a given $\lambda > 0$ asymptotically behaves as
\begin{equation}
\frac{N(\lambda)}{\sqrt{\lambda}}
\sim
\frac{1}{\pi} \int_{-L/2}^{L/2} \dd x\, \sqrt{\frac{w(x)}{p(x)}}
\end{equation}
for large $\lambda$; see, e.g., \cite{AtkinsonMingarelli:1987, NiessenZettl:1992}.
This holds under certain technical conditions on the functions $w(x)$, $p(x)$, and $q(x)$ in Eq.~\eqref{SL_op}, which are satisfied assuming our $v(x)$ and $K(x)$ are positive and smooth functions [if $q(x)$ is nonzero, then it is also assumed to be smooth].
Using Eqs.~\eqref{wpq_defs} and~\eqref{E_n} as well as Property~\ref{Item:SL1} in the main text, it follows that
\begin{equation}
\frac{n}{E_{n}}
\sim
\frac{1}{\pi} \int_{-L/2}^{L/2} \frac{\dd x}{v(x)}
= \frac{L}{\pi v_{0}} 
\end{equation}
for large $n$, where in the last step we used Eq.~\eqref{yx_v0_def}.
Thus, assuming that the integral defining $1/v_{0}$ converges, this gives the stated asymptotic behavior, $E_{n} = \pi v_{0} n/L + o(n)$, in the main text.
(Another possibility would be if the integral diverges, in which case $E_{n} \sim \textrm{const} \times n^{\gamma}$ for $\gamma < 1$, but this is inconsistent with that the eigenenergies must be proportional to integers $m_{n}$ with a common proportionality constant, in order for the theory to exhibit PWT.)

\vspace{-3mm}
\subsection*{Exact Bohr-Sommerfeld/WKB quantization condition}
\vspace{-3mm}

A stronger version of Weyl's law for our class of operators can be deduced from an exact generalization of the usual formulation of the Bohr-Sommerfeld/WKB quantization condition.
To state this condition, we consider our Sturm-Liouville problem written on Liouville normal form \eqref{SL_problem_mfu}:
\begin{equation}
\bigl[ - \partial_{y}^2 + V(y) \bigr] \mfu(y) = \Lambda \mfu(y)
\end{equation}
on the interval $I \equiv [-L/2, L/2] \ni y$, where we introduced $V(y) \equiv \hat{q}(y) / v_{0}^2$ and $\Lambda \equiv \lambda/v_{0}^2$ and recall that we impose Neumann BCs.
The first terms of the exact quantization condition for the $n$th eigenvalue $\Lambda_{n}$ can then be stated as
\begin{equation}
\label{eqc}
\phi(\Lambda_{n}) = \pi n
\end{equation}
with
\begin{equation}
\label{phi_def}
\phi(\Lambda) = \phi_{0}(\Lambda) + \phi_{1}(\Lambda) + O(\Lambda^{-5/2}), \vspace{-2.5mm}
\end{equation}
\begin{equation}
\label{phi0_phi1_def}
\phi_{0}(\Lambda)
\equiv \int_{I} \dd y\, \sqrt{\Lambda - V(y)},
\qquad
\phi_{1}(\Lambda)
\equiv - \frac{1}{4} \frac{\dd}{\dd \Lambda} \int_{I} \dd y\, \frac{V''(y)}{\sqrt{\Lambda - V(y)}}.
\end{equation}
This follows from \cite{Dunham:1932, BenderEtAl:1977, Frasca:2007} with the difference that we have no soft turning points but instead a finite interval with Neumann BCs: 
There is thus no Maslov index, and so the right-hand side of Eq.~\eqref{eqc} is $\pi n$ instead of $\pi (n + 1/2)$; cf.\ Sec.~2.4 in \cite{BrackBhaduri:2003}.

First, we consider $\phi_{1}(\Lambda)$.
Using the expansion $1/\sqrt{\Lambda - V(y)} = \Lambda^{-1/2} \bigl[ 1 + V(y)/ 2\Lambda + O(\Lambda^{-2}) \bigr]$ for large $\Lambda$, we obtain
\begin{equation}
\int_{I} \dd y\, \frac{V''(y)}{\sqrt{\Lambda - V(y)}}
= \frac{1}{\sqrt{\Lambda}} \int_{I} \dd y\, V''(y) + O(\Lambda^{-3/2}),
\end{equation}
which implies that
\begin{equation}
\phi_{1}(\Lambda)
= \frac{L}{8 \Lambda^{3/2}} \Delta + O(\Lambda^{-5/2})
\end{equation}
with
\begin{equation}
\label{Delta_def}
\Delta \equiv \frac{V'(L/2) - V'(-L/2)}{L}.
\end{equation}
Recall that $V(y)$ must be an even function on $I$ (for the system to exhibit PWT) and thus $V'(y)$ will be odd.
Second, we consider $\phi_{0}(\Lambda)$.
Using the expansion $\sqrt{\Lambda - V(y)} = \Lambda^{1/2} \bigl[ 1 - V(y)/ 2\Lambda - V(y)^2/8\Lambda^2 + O(\Lambda^{-3}) \bigr]$ for large $\Lambda$, we find
\begin{equation}
\phi_{0}(\Lambda)
= \sqrt{\Lambda} L \left( 1 - \frac{1}{2\Lambda} a_{1} - \frac{1}{8\Lambda^2} a_{2} \right) + O(\Lambda^{-5/2})
\end{equation}
with
\begin{equation}
\label{a1_a2}
a_1
\equiv \frac{1}{L} \int_{I} \dd y\, V(y),
\qquad
a_2
\equiv \frac{1}{L} \int_{I} \dd y\, V(y)^2,
\end{equation}
which can be interpreted as the first two moments of the potential $V(y)$.

Inserting the above into Eqs.~\eqref{eqc}--\eqref{phi0_phi1_def} yields
\begin{equation}
\sqrt{\Lambda_{n}} \left( 1 - \frac{1}{2\Lambda_{n}} a_{1} - \frac{1}{8\Lambda_{n}^2} [a_{2} - \Delta] + O(\Lambda_{n}^{-3}) \right) = \frac{\pi n}{L}.
\end{equation}
Since Eq.~\eqref{E_n} implies $\Lambda_{n} = \lambda_{n} / v_{0}^2 = ( E_{n} / v_{0} )^2$ and since $E_{n} = \pi m_{n}/T$ for integers $m_{n}$ satisfying the conditions~\ref{Item:mn_TLL_a}--\ref{Item:mn_TLL_b} on page~\pageref{Item:mn_TLL_a} of the main text, the quantization condition becomes
\begin{equation}
\label{eqc_mn}
m_{n} \left( 1 - \frac{L^2}{2 \pi^2 m_{n}^2} a_{1} - \frac{L^4}{8 \pi^4 m_{n}^4} [a_{2} - \Delta] + O(m_{n}^{-6}) \right) = n,
\end{equation}
where we used that $v_{0} = L/T$ by definition of $T$ as the common shortest possible time (for all modes) for the system to exhibit PWT.
Considering the difference $m_{n+1} - m_{n}$, it follows from Eq.~\eqref{eqc_mn} that $m_{n+1} - m_{n} = 1$ for all $n > n_{0}$ for some sufficiently large positive integer $n_0$.
The solution of this recursion formula is $m_n = n + a$ for some $a \in \mathbb{R}$ for $n > n_{0}$.
In other words, $m_n = n \bigl[ 1 + O(n^{-1})\bigr]$ for large $n$, which compared with Eq.~\eqref{eqc_mn} implies
\begin{equation}
a_{1} = 0,
\qquad
a_{2} - \Delta = 0.
\end{equation}
I.e., the potential $V(y)$ must have a vanishing mean, which implies that $a_2$ is the variance, which must equal $\Delta$.
[Note that Eqs.~\eqref{Delta_def} and~\eqref{a1_a2} then imply that $V'(L/2) \geq 0$, since $\Delta = a_2 \geq 0$ could otherwise not hold.]

The interval $I$ with Neumann BCs can be unfolded to $[-L/2, 3L/2]$ of length $2L$ with periodic BCs (similar to Part~A), which in turn can be viewed as the circle $S^1_{2L}$.
If we impose that $V(y)$ must be regular not only before but also after this unfolding procedure, then the only possibility is $V'(L/2) = 0$, since $V(y)$ must be even, from which it follows that $\Delta = 0$ in Eq.~\eqref{Delta_def}.
Consequently, both the mean $a_1$ and the variance $a_2$ of the potential $V(y)$ vanish in that case, which then implies that $\hat{q}(y) = v_0^2 V(y)$ must be zero (almost everywhere).
From Eq.~\eqref{qhat} with $q(x) = 0$ and reflection symmetry, it follows that $K(x)$ must be constant for PWT to be exhibited in an inhomogeneous TLL with the assumption of regularity after unfolding to $S^1_{2L}$.
However, if such regularity is not assumed, then the quantization condition \eqref{eqc_mn} does not prevent $\hat{q}(y) = v_0^2 V(y)$ from being nonzero, which in turn corresponds to a nonconstant $K(x)$.

\vspace{-3mm}
\subsection*{``Massive'' inhomogeneous TLLs}
\vspace{-3mm}

For massless TLL bosons, as discussed above and in the main text, given a spectrum satisfying the PWT requirements, we saw that $K(x)$ can be found uniquely up to rescaling and changing coordinates.

We now extend to SL operators \eqref{SL_op} with a smooth $q(x) \neq 0$, corresponding to an inhomogeneous (possibly imaginary) mass $M(x) = \sqrt{-q(x)/v_{0}^2 v(x)}$ by adding $v_{0}^2 M(x)^2 \varphi(x)^2$ inside the big parentheses in Eq.~\eqref{H_iTLL}:
\begin{equation}
\label{H_miTLL}
H_{\textrm{miTLL}}
\equiv
\int_{-L/2}^{L/2} \frac{\dd x}{2\pi} v(x)
  \! \wick{
    \biggl(
      \frac{[\pi \Pi(x)]^2}{K(x)}
      + K(x)[\partial_x \varphi(x)]^2
      + v_{0}^2 M(x)^2 \varphi(x)^2
    \biggr)
  } \! .
\end{equation}
Indeed, it is straightforward to extend the previous discussion to such models.
For instance, the PWT requirements on the spectrum are the same, except the spectrum does not start at zero:
For $q(x) \neq 0$, a constant $u_{0}(x)$ no longer solves Eq.~\eqref{SL_problem}, thus $E_{0} > 0$.
Instead, we have an inhomogeneous TLL with a mass term, for which one can show the following statement:
\emph{``Massive'' inhomogeneous TLLs exhibit PWT as defined in Eq.~\eqref{PWT_def} if and only if $v(x)$, $K(x)$, and $q(x)$ are even and the eigenenergies are $E_{n} = \pi \tilde{m}_{n}/T$ for integers $\tilde{m}_{n}$ so that}
\begin{enumerate}[label = \textnormal{(\alph*)}, topsep=0.75ex, itemsep=0.0ex, partopsep=1ex, parsep=0.75ex]

\item
\label{Item:mn_mTLL_a}
\emph{$0 \leq \tilde{m}_{0} < \tilde{m}_{1} < \ldots$ with each $\tilde{m}_{n}$ even or odd if $n$ is even or odd,}

\item
\label{Item:mn_mTLL_b}
\emph{$\tilde{m}_{n} \sim n$ for large $n$, i.e., $\tilde{m}_{n} = n + o(n)$.}

\end{enumerate}

The integers in the result above can be written as $\tilde{m}_{n} = m_{n} + k$ in terms of the integers $m_{n}$ for the massless case in the main text, allowing for an overall shift by a finite even integer $k$.
[The latter must be even so that \ref{Item:mn_mTLL_a} above is satisfied, as required due to Property~\ref{Item:SL2} in the main text.]
Now, given such a spectrum, the results in \cite{Borg:1946, Levinson:1949} still imply that $\hat{q}(y)$ is uniquely determined by it, but this can no longer uniquely determine $K(x)$ since the first term on the right-hand side of Eq.~\eqref{qhat} is nonzero in general.

As an example, we revisit the case with $v(x) = v \sqrt{1 - (2x/L)^2}$ and $K(x) = K \bigl[ 1 - (2x/L)^2 \bigr]^\alpha$ for $\alpha > -1/2$ and constants $v, K > 0$ encountered in the main text, whose eigenfunctions were given by Gegenbauer polynomials $C^{(\alpha)}_{n}(2x/L)$, but now we include a fine-tuned potential $q(x) = - (2v\alpha/L)^2 w(x)$, where we recall that $w(x) = K(x)/v(x)$.
This fine-tuned potential allows us to complete the square so that the eigenvalues in Eq.~\eqref{SL_problem} are $\lambda_{n} = (\pi v_{0}/L) (n + \alpha)^2$  with $v_{0} = 2v/\pi$, meaning that $E_{n} = \pi (n + \alpha) v_{0}/L$.
Consequently, any $\alpha = k = 0, 2, 4, \ldots$ works, yielding a discrete family of models with fine-tuned inhomogeneous mass that exhibit PWT.
We stress that the mass in this example satisfies $M(x) \propto 1/L$ and therefore vanishes in the thermodynamic limit.
Thus, one should not expect hallmarks of massive theories, such as exponential decay of two-point correlations with respect to the spatial separation.

From the theory of regular SL operators and the Bohr-Sommerfeld/WKB quantization condition in the previous section, if one assumes regularity also after unfolding to $S^1_{2L}$, then ``massive'' inhomogeneous TLLs must lie in the orbit of $\hat{q}(y) = 0$ in order to exhibit PWT.
I.e., in that case, while $K(x)$ does not have to be constant, the choice of $q(x)$ must be such that $\hat{q}(y)$ given by Eq.~\eqref{qhat} is still zero, corresponding to an exactly linear spectrum as stated in the main text.
However, the above Gegenbauer example is irregular, meaning that it does not have to lie in the orbit of $\hat{q}(y) = 0$ (which it does not for $\alpha = 2, 4, \ldots$) while exhibiting PWT.

\vspace{-3mm}
\subsection*{Liouville normal form and compactified free bosons}
\vspace{-3mm}

It is curious to note the following ``shortcut'' to writing our massless boson model \eqref{H_iTLL} in Liouville normal form:
Recall that inhomogeneous TLL theory is that of compactified free bosons in curved spacetime with a position-dependent compactification radius $R(x)$ and the metric $(h_{\mu\nu}) = \diag(v(x)^2/v_{0}^2, -1)$ in coordinates $(x^0, x^1) \equiv (v_{0}t, x)$; see, e.g., \cite{GMS:2022, Moosavi:DBdG_iQLs:2023}.
Given a Luttinger parameter profile $K(x)$, the compactification radius is $R(x) \equiv \sqrt{2\alpha'} K(x)$ ($\alpha'$ has dimensions of length squared and is commonly set equal to $2$).
In terms of the bosonic field $\varphi(x)$ (with values taken modulo $2\pi$), the action corresponding to the Hamiltonian \eqref{H_iTLL} is
\begin{equation}
S_{\textrm{iTLL}}
\equiv
\frac{1}{4\pi\alpha'} \int \dd^2 x\, \sqrt{-h}
R(x)^2 (\partial_{\mu} \varphi) (\partial^{\mu} \varphi),
\end{equation}
where we recall that $h \equiv \det(h_{\mu\nu})$.
This can be viewed as a nonexactly marginal deformation of a theory with constant compactification radius; cf.\ \cite{Moosavi:DBdG_iQLs:2023}.
The corresponding action for $X(x) \equiv R(x) \varphi(x)$ becomes
\begin{equation}
S_{\textrm{iTLL}}
= \frac{1}{4\pi \alpha'} \int \dd^2 x\, \sqrt{-h} \bigl[ (\partial_{\mu} X) (\partial^{\mu} X) - \hat{q} X^2/v(x)^2 \bigr]
\end{equation}
with $\hat{q} = \hat{q}(y)$ in Eq.~\eqref{qhat} for $y = y(x)$ given by Eq.~\eqref{yx_v0_def}, which agrees with Eq.~\eqref{SL_problem_mfu} for $\mfu(y) = X(x)/\sqrt{2\alpha'}$.
I.e., as a theory in terms of the field $X(x)$, even when $q(x) = 0$, the position-dependent $K(x)$ corresponds a (possibly imaginary) inhomogeneous mass $\sqrt{\hat{q}(y)}/v(x)$.
This also extends to the case
\begin{equation}
S_{\textrm{miTLL}}
\equiv
\frac{1}{2\pi} \int \dd^2 x \sqrt{-h}
\bigl[ K(x) (\partial_{\mu} \varphi) (\partial^{\mu} \varphi) - v_{0}^2 M(x)^2 \varphi^2 \bigr]
\end{equation}
with a nonzero mass $M(x) = \sqrt{-q(x)/v_{0}^2 v(x)}$ for the bosonic field $\varphi(x)$, corresponding to the Hamiltonian \eqref{H_miTLL}, the only difference being that the first term on the right-hand side of Eq.~\eqref{qhat} is then no longer zero.

\vspace{-4mm}
\section*{Part~C: Correlation functions in inhomogeneous TLLs}
\vspace{-3mm}

Consider the inhomogeneous TLL theory with Hamiltonian $H_{\textrm{iTLL}}$ in Eq.~\eqref{H_iTLL}.
Using the eigenfunctions obtained from SL theory, we can compute any correlation function for the fields $\varphi(x)$ and $\vartheta(x)$, where the latter is defined by
\begin{equation}
\partial_x \vartheta(x)
\equiv
\pi \Pi(x).
\end{equation}
Case in point, using the expansions in Eq.~\eqref{varphi_Pi_expansions_time} with zero modes omitted, that $a_{n}$ annihilates the ground state $|\Omega\rangle$ of $H_{\textrm{iTLL}}$, and the commutation relations $\bigl[ a\pdag_{n\ppr}, a^\dagger_{n'} \bigr] = \delta_{n,n'}$, it is straightforward to show that the ground-state two-point correlations are
\begin{equation}
\begin{aligned}
\langle\Omega| \varphi(x,t) \varphi(x',t') |\Omega\rangle
& = \sum_{n = 1}^{\infty} \frac{\pi}{2E_{n}} u_{n}(x)u_{n}(x') \ee^{-\ii E_{n} (t-t')}, \\
\langle\Omega| \varphi(x,t) \vartheta(x',t') |\Omega\rangle
& = \ii \sum_{n = 1}^{\infty} \frac{\pi}{2} u_{n}(x)U_{n}(x') \ee^{-\ii E_{n} (t-t')}, \\
\langle\Omega| \vartheta(x,t) \vartheta(x',t') |\Omega\rangle
& = \sum_{n = 1}^{\infty} \frac{\pi E_{n}}{2} U_{n}(x)U_{n}(x') \ee^{-\ii E_{n} (t-t')},
\end{aligned}
\end{equation}
where we introduced $U_{n}(x)$ defined by
\begin{equation}
\partial_x U_{n}(x)
\equiv
w(x) u_{n}(x).
\end{equation}
The first of these correlations is the same as the one in Eq.~\eqref{varphivarphi} in the main text.
Similar correlations can be computed for other objects, including vertex operators, i.e., normal-ordered exponentials of the bosonic fields \cite{Voit:1995, DelftSchoeller:1998, SchulzCunibertiPieri:2000, Giamarchi:2003, Cazalilla:2004, LaMo1:2015}.
It is also straightforward to compute Green's functions; see, e.g., \cite{GMS:2022}, the only difference with the present case being that one must account for the nontrivial weight $w(x)$ that was simply a constant in \cite{GMS:2022}.

As an important example, consider a theory with $K(x) = K > 0$ constant but $v(x) > 0$ general [and $q(x) = 0$].
We recall that the SL operator in this case is $\cA = - v(x) \partial_{x} v(x) \partial_{x} = -v_{0}^2 \partial_{y}^2$ using $y = y(x)$ given by Eq.~\eqref{yx_v0_def} with eigenenergies $E_{n} = \pi n v_{0}/L$ and eigenfunctions $u_{n > 0}(x) = \sqrt{2 v_{0}/L} \cos \bigl( n\pi [L + 2y]/2L \bigr)$ and $u_{0}(x) = \sqrt{v_{0}/L}$.
One can then show that the expression \eqref{varphivarphi} for the ground-state $\varphi\varphi$ correlation function (with zero modes omitted) yields
\begin{multline}
\label{phiphi_FF}
\hspace{-2.2mm} \langle\Omega| \varphi(x,t) \varphi(x',t') |\Omega\rangle \\
\begin{aligned}
= \frac{1}{4}
& \left(
  \log
  \frac{
    \ii \exp \Bigl( -\ii\frac{\pi}{2L} [y(x)-y(x')-v_{0}(t-t')] \Bigr)
  }{
    2 \sin \Bigl( \frac{\pi}{2L} [y(x)-y(x')-v_{0}(t-t')+\ii0^+] \Bigr)
  }
+ \log
  \frac{
    \ii \exp \Bigl( -\ii\frac{\pi}{2L} [y(x)+y(x')+L-v_{0}(t-t')] \Bigr)
  }{
    2 \sin \Bigl( \frac{\pi}{2L} [y(x)+y(x')+L-v_{0}(t-t')+\ii0^+] \Bigr)
  }
  \right. \\
& \quad + \left.
  \log
  \frac{
    -\ii \exp \Bigl( \ii\frac{\pi}{2L} [y(x)+y(x')+L+v_{0}(t-t')] \Bigr)
  }{
    2 \sin \Bigl( \frac{\pi}{2L} [y(x)+y(x')+L+v_{0}(t-t')-\ii0^+] \Bigr)
  }
+ \log
  \frac{
    -\ii \exp \Bigl( \ii\frac{\pi}{2L} [y(x)-y(x')+v_{0}(t-t')] \Bigr)
  }{
    2 \sin \Bigl( \frac{\pi}{2L} [y(x)-y(x')+v_{0}(t-t')-\ii0^+] \Bigr)
  }
  \right),
\end{aligned}
\end{multline}
where we included necessary regularizations in order to interpret coincident spacetime points and used the identity $\sum_{n = 1}^{\infty} \frac{1}{n} \ee^{-\xi n} = - \log \left( 1 - \ee^{-\xi} \right)$ for any $\xi$ such that $\Re \xi > 0$.
The real part of Eq.~\eqref{phiphi_FF} gives the analytical expression for what was plotted in Fig.~\ref{Fig:MCGS_fig1} (assuming $K = 1$).
Moreover, for deltalike waves, the terms appearing in $\log F(t)$ given by Eq.~\eqref{Ft_final} for the free-fermion case \eqref{G_FF} closely resemble those in Eq.~\eqref{phiphi_FF}.
Indeed, the former agrees with $4 \langle\Omega| \varphi(x,t) \varphi(x',t') |\Omega\rangle$ to leading order, and one can furthermore show that the agreement between the two ways of computing (CFT methods versus bosonization) is exact (as it should be) by evaluating and summing all combinations of two-point correlation functions of the vertex operators describing the right- and left-moving free fermionic fields.


\bigskip

\noindent {\bf References}

{\small%
\begin{enumerate}[leftmargin=2.5em, itemsep=-0.1em, label={[\arabic*]}, ref={\arabic*}, itemindent=0.0em]

\item[{[34]}]
\label{GLM:2018_SM}
K.~Gaw\k{e}dzki, E.~Langmann, and P.~Moosavi,
Finite-time universality in nonequilibrium CFT,
\href{https://doi.org/10.1007/s10955-018-2025-x}{J.\ Stat.\ Phys.\ {\bf 172}, 353 (2018)}.

\item[{[36]}]
\label{Moosavi:iCFT:2024_SM}
P.~Moosavi,
Inhomogeneous conformal field theory out of equilibrium,
\href{https://doi.org/10.1007/s00023-021-01118-0}{Ann.\ Henri Poincar\'{e} {\bf 25}, 1083 (2024)}.

\item[{[37]}]
\label{GMS:2022_SM}
M.~Gluza, P.~Moosavi, and S.~Sotiriadis,
Breaking of Huygens-Fresnel principle in inhomogeneous Tomonaga-Luttinger liquids,
\href{https://doi.org/10.1088/1751-8121/ac39cc}{J.\ Phys.\ A: Math.\ Theor.\ {\bf 55}, 054002 (2022)}.

\item[{[48]}]
\label{GawedzkiKozlowski:2020_SM}
K.~Gaw\k{e}dzki and K.~K.~Kozlowski,
Full counting statistics of energy transfers in inhomogeneous nonequilibrium
states of (1+1)D CFT,
\href{https://doi.org/10.1007/s00220-020-03774-5}{Commun.\ Math.\ Phys.\ {\bf 377}, 1227 (2020)}.

\item[{[54]}]
\label{Tomonaga:1950_SM}
S.~Tomonaga,
Remarks on Bloch's method of sound waves applied to many-fermion problems, \href{https://doi.org/10.1143/ptp/5.4.544}{Prog.\ Theor.\ Phys.\ {\bf 5}, 544 (1950)}.

\item[{[55]}]
\label{Luttinger:1963_SM}
J.~M.~Luttinger,
An exactly soluble model of a many-fermion system,
\href{https://doi.org/10.1063/1.1704046}{J.\ Math.\ Phys.\ {\bf 4}, 1154 (1963)}.

\item[{[56]}]
\label{MattisLieb:1965_SM}
D.~C.~Mattis and E.~H.~Lieb,
Exact solution of a many-fermion system and its associated boson field,
\href{https://doi.org/10.1063/1.1704281}{J.\ Math.\ Phys.\ {\bf 6}, 304 (1965)}.

\item[{[61]}]
\label{Moosavi:DBdG_iQLs:2023_SM}
P.~Moosavi,
Exact Dirac-Bogoliubov-de Gennes dynamics for inhomogeneous quantum liquids,
\href{https://doi.org/10.1103/PhysRevLett.131.100401}{Phys.\ Rev.\ Lett.\ {\bf 131}, 100401 (2023)}.

\item[{[62]}]
\label{Voit:1995_SM}
J.~Voit,
One-dimensional Fermi liquids,
\href{https://doi.org/10.1088/0034-4885/58/9/002}{Rep.\ Prog.\ Phys.\ {\bf 58}, 977 (1995)}.

\item[{[63]}]
\label{DelftSchoeller:1998_SM}
J.~von~Delft and H.~Schoeller,
Bosonization for beginners---refermionization for experts,
\href{https://doi.org/10.1002/andp.19985100401}{Ann.\ Phys.\ {\bf 7}, 225 (1998)}.

\item[{[64]}]
\label{SchulzCunibertiPieri:2000_SM}
H.~J.~Schulz, G.~Cuniberti, and P.~Pieri,
Fermi liquids and Luttinger liquids,
in \href{https://doi.org/10.1007/978-3-662-04273-1}{{\it Field Theories for Low-Dimensional Condensed Matter Systems}}, Springer Series in Solid-State Sciences Vol.\ 131, edited by G.~Morandi, P.~Sodano, A.~Tagliacozzo, and V.~Tognett
(Springer, Berlin, Heidelberg, 2000), p.\ 9,
\href{https://arxiv.org/abs/cond-mat/9807366}{arXiv:cond-mat/9807366}.

\item[{[65]}]
\label{Giamarchi:2003_SM}
T.~Giamarchi,
\href{https://doi.org/10.1093/acprof:oso/9780198525004.001.0001}{{\it Quantum Physics in One Dimension}}
(Oxford University Press, Oxford, 2003).

\item[{[66]}]
\label{Cazalilla:2004_SM}
M.~A.~Cazalilla,
Bosonizing one-dimensional cold atomic gases,
\href{https://doi.org/10.1088/0953-4075/37/7/051}{J.\ Phys.\ B: At.\ Mol.\ Opt.\ Phys.\ {\bf 37}, S1 (2004)}.

\item[{[67]}]
\label{LaMo1:2015_SM}
E.~Langmann and P.~Moosavi,
Construction by bosonization of a fermion-phonon model, 
\href{https://doi.org/10.1063/1.4930299}{J.\ Math.\ Phys.\ {\bf 56}, 091902 (2015)}.

\item[{[75]}]
\label{EggertAffleck:1992_SM}
S.~Eggert and I.~Affleck,
Magnetic impurities in half-integer-spin Heisenberg antiferromagnetic chains, \href{https://doi.org/10.1103/PhysRevB.46.10866}{Phys.\ Rev.\ B {\bf 46}, 10866 (1992)}.

\item[{[76]}]
\label{FabrizioGogolin:1995_SM}
M.~Fabrizio and A.~O.~Gogolin,
Interacting one-dimensional electron gas with open boundaries,
\href{https://doi.org/10.1103/PhysRevB.51.17827}{Phys.\ Rev.\ B {\bf 51}, 17827 (1995)}.

\item[{[77]}]
\label{ChuaEtAl:2020_SM}
V.~Chua, K.~Laubscher, J.~Klinovaja, and D.~Loss,
Majorana zero modes and their bosonization,
\href{https://doi.org/10.1103/PhysRevB.102.155416}{Phys.\ Rev.\ B {\bf 102}, 155416 (2020)}.

\item[{[97]}]
\label{AtkinsonMingarelli:1987_SM}
F.~V.~Atkinson and A.~B.~Mingarelli,
Asymptotics of the number of zeros and of the eigenvalues of general weighted Sturm-Liouville problems,
\href{https://doi.org/10.1515/crll.1987.375-376.380}{J.\ Reine Angew.\ Math.\ {\bf 1987}, 380 (1987)}.

\item[{[98]}]
\label{NiessenZettl:1992_SM}
H.-D.~Niessen and A.~Zettl,
Singular Sturm-Liouville problems: The Friedrichs extension and comparison of eigenvalues,
\href{https://doi.org/10.1112/plms/s3-64.3.545}{Proc.\ London Math.\ Soc.\ {\bf s3-64}, 545 (1992)}.

\item[{[105]}]
\label{Borg:1946_SM}
G.~Borg,
Eine Umkehrung der Sturm-Liouvilleschen Eigenwertaufgabe,
\href{https://doi.org/10.1007/BF02421600}{Acta Math.\ {\bf 78}, 1 (1946)}.

\item[{[106]}]
\label{Levinson:1949_SM}
N.~Levinson,
The inverse Sturm-Liouville problem,
\href{http://www.jstor.org/stable/24527827}{Mat.\ Tidsskr.\ B {\bf 1949}, 25 (1949)}.

\item[{[110]}]
\label{Dunham:1932_SM}
J.~L.~Dunham,
The Wentzel-Brillouin-Kramers method
of solving the wave equation,
\href{https://doi.org/10.1103/PhysRev.41.713}{Phys.\ Rev.\ {\bf 41}, 713 (1932)}.

\item[{[111]}]
\label{BenderEtAl:1977_SM}
C.~M.~Bender, K.~Olaussen, and P.~S.~Wang, 
Numerological analysis of the WKB approximation in large order,
\href{https://doi.org/10.1103/PhysRevD.16.1740}{Phys.\ Rev.\ D {\bf 16}, 1740 (1977)}.

\item[{[112]}]
\label{Frasca:2007_SM}
M.~Frasca,
A strongly perturbed quantum system is a semiclassical system,
\href{https://doi.org/10.1098/rspa.2007.1879}{Proc.\ R.\ Soc.\ A {\bf 463}, 2195 (2007)}.

\item[{[113]}]
\label{BrackBhaduri:2003_SM}
M.~Brack and R.~Bhaduri,
\href{https://doi.org/10.1201/9780429502828}{{\it Semiclassical Physics}}
(CRC Press, Boca Raton, 2003).

\end{enumerate}
}


\begin{thebibliography}{119}
\bibitem [{Not({\natexlab{b}})}]{Note:Particle_flow}%
  \BibitemOpen
  \bibinfo {note} {{This implies zero
  particle flow at $x = \pm L/2$ since the two fermion densities $\rho_{\pm}(x)
  = \wick{ \psi^\dagger_{\pm}(x)\psi\pdag_{\pm}(x) }$ are equal at each
  boundary and the current is $\propto \rho_{+} - \rho_{-}$.}}\BibitemShut
  {Stop}%
\bibitem [{Not({\natexlab{c}})}]{Note:Eq_psi_def}%
  \BibitemOpen
  \bibinfo {note} {{Eq.~\eqref{psi_def}
  corresponds to the low-energy linearization of the spin-chain dispersion
  relation around its Fermi points $\pm \kF$, followed by recentering the
  resulting fields so that they respectively describe right- and left-moving excitations around
  zero momentum.}}\BibitemShut {Stop}%
\bibitem [{Not({\natexlab{e}})}]{Note:Limit_Mobius}%
  \BibitemOpen
  \bibinfo {note} {{Oppositely, the
  limiting case of an inhomogeneous CFT given by M{\"o}bius transformations so
  that $v(x)$ vanishes at $\pm L/2$ cannot exhibit PWT since $1/v_{0}$
  diverges.}}\BibitemShut {Stop}%
\bibitem [{Not({\natexlab{f}})}]{Note:BCs_K(x)}%
  \BibitemOpen
  \bibinfo {note} {{This applies to regular
  systems, to ensure that the Neumann BCs remain simple. If irregular, say,
  $K(x)$ and/or $v(x)$ vanishing at $\pm L/2$, this is irrelevant as the BCs
  are different and no longer couple right and left movers.}}\BibitemShut {Stop}%
\bibitem [{Not({\natexlab{g}})}]{Note:GMS_vs_DBdG}%
  \BibitemOpen
  \bibinfo {note} {{This generalizes
  \cite{GMS:2022} as $v(x)$ and $K(x)$ are unrelated, but in a way that is different from
  the approach in \cite{Moosavi:DBdG_iQLs:2023}.}}\BibitemShut {Stop}%
\bibitem [{Not({\natexlab{i}})}]{Note:Reg_vs_irreg}%
  \BibitemOpen
  \bibinfo {note} {{Here, these amount to
  $v(x)$ and $K(x)$ being strictly positive and $v(x)$, $K(x)$, $\partial_x p(x)$
  [and $q(x)$ if nonzero] continuous on the whole interval $[-L/2,
  L/2]$.}}\BibitemShut {Stop}%
\bibitem [{Not({\natexlab{j}})}]{Note:Zero_modes}%
  \BibitemOpen
  \bibinfo {note} {{Zero modes such as
  $\varphi_{0}$ need separate treatments when $E_{0} = 0$, but they have no
  consequence for our discussion of PWT, so we omit them.}}\BibitemShut
  {Stop}%
\bibitem [{Not({\natexlab{l}})}]{Note:GLM_eq}%
  \BibitemOpen
  \bibinfo {note} {{A related prominent
  problem is the inverse scattering for integrable differential equations,
  solved by the Gelfand-Levitan-Marchenko equation. That case is on the
  infinite line, while we have a finite interval.}}\BibitemShut {Stop}%
\bibitem [{Not({\natexlab{m}})}]{Note:Irreg_BCs}%
  \BibitemOpen
  \bibinfo {note} {{The BCs are that
  solutions must be sufficiently well behaved for Eq.~\eqref{jmath_BCs} to be
  satisfied.}}\BibitemShut {Stop}%
\bibitem [{Not({\natexlab{n}})}]{Note:Unfolding_K(x)}%
  \BibitemOpen
  \bibinfo {note} {{As $K(x)$ must be even,
  this also implies $K'(x)\big|_{x = \pm L/2} = 0$, as assumed below
  Eq.~\eqref{jmath_BCs}.}}\BibitemShut {Stop}%
\end{thebibliography}
\end{document}